\documentclass[12pt,draftcls,onecolumn]{IEEEtran}

\newcommand{\e}{{\rm e}}
\usepackage{amsfonts}
\usepackage{color}
\usepackage{amsmath}
\usepackage{amsthm}
\usepackage{amssymb}
\usepackage{t1enc}
\usepackage{alltt}
\usepackage{epsfig}
\usepackage{mathrsfs}
\usepackage{graphicx,ifthen}
\usepackage{multirow}
\usepackage{subfigure}
\usepackage{hyperref}

\newtheorem{proposition}{Proposition}
\makeatletter
\newcommand{\rmnum}[1]{\romannumeral #1}
\newcommand{\Rmnum}[1]{\expandafter\@slowromancap\romannumeral #1@}

\begin{document}

\title{Spectrum Sensing in the Presence of Multiple Primary Users}

\author{Lu~Wei,~\IEEEmembership{Student~Member,~IEEE,} and~Olav~Tirkkonen,~\IEEEmembership{Member,~IEEE}%
\thanks{The authors are with the Department of Communications and Networking, Aalto University, Finland (E-mail: \{lu.wei,
olav.tirkkonen\}@aalto.fi). This work is partially supported by the Academy of Finland
under the project Spectrum Management for Future Wireless
Systems (Grant No.: $133652$).}%
\thanks{Partial result of this paper, i.e. Proposition $1$, was presented in the $6$th International Conference on Cognitive Radio Oriented Wireless Networks and Communications (CrownCom), Jun. $2011$.}%
}

\markboth{Accepted in IEEE Transactions on Communications}%
{Lu \MakeLowercase{\textit{et al.}}: Cooperative Spectrum Sensing in
the Presence of Multiple Primary Users}

\maketitle

\begin{abstract}

We consider multi-antenna cooperative spectrum sensing in cognitive radio networks, when there may be multiple primary users. A detector based on the spherical test is analyzed in such a scenario. Based on the moments of the distributions involved, simple and accurate analytical formulae for the key performance metrics of the detector are derived. The false alarm and the detection probabilities, as well as the detection threshold and Receiver Operation Characteristics are available in closed form. Simulations are provided to verify the accuracy of the derived results, and to compare with other detectors in realistic sensing scenarios.

\end{abstract}
\


\IEEEpeerreviewmaketitle

\section{Introduction}

Cognitive radio (CR) is a promising technique for future wireless
communication systems. In CR networks, dynamic spectrum access is
implemented to mitigate spectrum scarcity.  A secondary (unlicensed)
user is allowed to utilize the spectrum resources when it does not
cause intolerable interference to the primary (licensed) user. A key
requirement for this is the secondary user's ability to detect the
presence of the primary user. Thus spectrum sensing is considered as
a key component in CR networks.

Prior work on cooperative spectrum sensing predominately employ the
assumption of a single active primary user. Based on this
assumption, several eigenvalue based sensing algorithms have been
proposed
recently~\cite{2008Yonghong,2009Kritchman,2010Taherpour,2009Lu,2008bZeng,2010Wang,2010Bianchi,2011Nadler}.
These algorithms are non-parametric, i.e. they do not require
information of the primary user, in contrast to e.g. feature
detection. Also, they achieve optimality under different assumptions on the knowledge of the parameters. The assumption of a single primary user is made as the
investigations in the literature have mainly focussed on CR
networks, where the primary users are TV or DVB systems. In these
systems the single active primary user assumption is, to some
extent, justifiable. In addition, assuming a single primary user
leads to analytically tractable problems.

The single primary user assumption may fail to reflect the situation
in forthcoming CR networks, where the primary system could be a
cellular network, and the existence of more than one primary user
would be the prevailing condition. Using existing single primary
user detection algorithms in such a scenario will induce performance
loss. Despite the need to understand multiple primary user
detection, the results in this direction are rather limited. A
heuristic detection algorithm based on the ratio of the extreme eigenvalues is investigated
in~\cite{2008Zeng,2009Federico,2010Federico}, but its detection
performance turns out to be sub-optimal~\cite{2011Nadler}. Recently,
a novel detection algorithm in the presence of multiple primary
users, based on the spherical test, has been proposed in~\cite{2010Zhang}. However, no analytical
results pertaining to its statistical performance were presented. In this paper we analytically investigate the detection performance by deriving closed-form approximations for the test statistics distributions under both hypothesis. These approximations are obtained by matching the moments of the test statistics to the Beta distribution. Using the derived results we obtain analytical formulae for major performance measures, such as the false alarm probability, the detection probability, the decision threshold and the Receiver Operating Characteristic (ROC). The derived approximations are easily computable and simulations show that they are accurate for the considered sensor sizes, number of samples, the assumed number of primary users and corresponding SNRs. In addition, for the most useful system configuration of two sensors with arbitrary number of samples a simple form of the exact detection probability is derived.

The rest of this paper is organized as follows. In
Section~\ref{sec:Formulation} we study the test statistics for the
multiple primary user detection after outlining the signal model.
Performance analysis of the chosen detection algorithm is
addressed in Section~\ref{sec:Distributions}.
Section~\ref{sec:Simulations} presents numerical examples to examine
the detection performance in diverse scenarios. Finally in
Section~\ref{sec:Conclusion} we conclude the main results of this
paper and point out some possible future research directions based on the results of this work.

\section{Problem Formulation}\label{sec:Formulation}

\subsection{Signal Model}

Consider the standard model for $K$-sensor cooperative detection in
the presence of $P$ primary users,
\begin{equation}\label{eq:model}
\mathbf{x}=\mathbf{Hs}+\sigma\mathbf{n}
\end{equation}
where $\mathbf{x}\in\mathbb{C}^{K}$ is the received data vector. The
$K$ sensors may be e.g. $K$ receive antennas in one secondary
terminal or $K$ secondary devices each with a single antenna, or any
combination of these. The $K\times P$ matrix
$\mathbf{H}=[\mathbf{h}_{1},\ldots,\mathbf{h}_{P}]$ represents the
channels between the $P$ primary users and the $K$ sensors. The
$P\times 1$ vector $\mathbf{s}=[s_{1},\ldots,s_{P}]'$ denotes zero mean transmitted signals
from the primary users. The $K\times 1$ vector $\sigma\mathbf{n}$ is the
complex Gaussian noise with zero mean and covariance matrix
$\sigma^{2}\mathbf{I}_{K}$, where the scalar $\sigma^{2}$ is the noise power.

We collect $N$ i.i.d observations from model~(\ref{eq:model}) to a
$K\times N$ matrix $\mathbf{X} =
[\mathbf{x}_{1},\ldots,\mathbf{x}_{N}]$. The problem of interest is
to use the data matrix $\mathbf{X}$ to decide whether there are
primary users.\footnote{This collaborative sensing scenario is more relevant when the $K$ sensors are in one device, since for multiple collaborating devices, accurate time synchronization between devices are needed and communications to the fusion center becomes an issue. Typically, $K$ is less than eight due to physical constraints of the device size.} For ease of analysis we make the following assumptions
\begin{enumerate}
\item The channel $\mathbf{H}$ is constant during sensing time.
\item The primary user's signal follows an i.i.d zero mean Gaussian distribution and is uncorrelated with the noise.
\end{enumerate}
Due to the first assumption the channel model for $\mathbf{H}$ may
not need to be specified. In the absence of primary users, the
sample covariance matrix $\mathbf{R}=\mathbf{XX^{\dag}}$ follows an
uncorrelated (white) complex Wishart distribution
$\mathcal{W}_{K}\left(N,\mathbf{\Sigma}\right)$ with population
covariance matrix
\begin{equation}\label{eq:coH0}
\mathbf{\Sigma}:=\mathbb{E}[\mathbf{XX^{\dag}}]/N=\sigma^{2}\mathbf{I}_{K}.
\end{equation}
In the presence of primary users, by the two assumptions above, the sample covariance matrix
$\mathbf{R}$ follows a correlated complex Wishart distribution. The correlation is given by the presence of the signals with the
covariance matrix equals
\begin{equation}\label{eq:coH1}
\mathbf{\Sigma}=\sigma^{2}\mathbf{I}_{K}+\sum_{i=1}^{P}\gamma_{i}\mathbf{h}_{i}\mathbf{h}^{\dag}_{i},
\end{equation}
where $\gamma_{i}:=\mathbb{E}[s_{i}s_{i}^{\dag}]$ defines the transmission power of the $i$-th primary
user. The received Signal to Noise Ratio (SNR) of primary user $i$
across the $K$ sensors is
\begin{equation}
\text{SNR}_{i}:=\frac{\gamma_{i}||\mathbf{h}_{i}||^{2}}{\sigma^{2}}.
\end{equation}
Finally, we denote the ordered eigenvalues of the sample covariance
matrix $\mathbf{R}$ by $\lambda_1 \geq \lambda_2 \geq \ldots \geq
\lambda_K$.

\subsection{Test Statistics}

The differences between the population covariance
matrices~(\ref{eq:coH0}) and~(\ref{eq:coH1}) can be explored to
detect the primary user. This detection problem can be formulated as
a binary hypothesis test, where hypothesis $\mathcal{H}_{0}$ denotes
the absence of primary users and hypothesis $\mathcal{H}_{1}$
denotes the presence of primary users. Declaring wrongly
$\mathcal{H}_{0}$, or declaring correctly $\mathcal{H}_{1}$, defines the false alarm
probability $P_{\text{fa}}$, and the detection probability
$P_{\text{d}}$, respectively.
Since the sample covariance matrix $\mathbf{R}$ is a Wishart matrix, it is sufficient statistics
for the population covariance matrix
$\mathbf{\Sigma}$~\cite{2003Anderson}. This leads to various test
statistics as functions of $\mathbf{R}$ with different assumptions
on the number of primary users $P$, and the knowledge of the noise
power $\sigma^{2}$.

In the case of \emph{a single} primary user ($P=1$) the hypothesis
test can be expressed as
\begin{eqnarray}
\mathcal{H}_{0}&:& \mathbf{\Sigma}=\sigma^{2}\mathbf{I}_{K}\\
\mathcal{H}_{1}&:&
\mathbf{\Sigma}=\sigma^{2}\mathbf{I}_{K}+\gamma_{1}\mathbf{h}_{1}\mathbf{h}^{\dag}_{1}.
\end{eqnarray}
Assuming known noise power $\sigma^{2}$ the Largest Eigenvalue based (LE) detection $\left(T_{\text{LE}}=\lambda_{1}\right)$ is shown to be optimal under the Generalized Likelihood Ratio Test (GLRT) criterion~\cite{2010Taherpour}. Performance analysis of the LE detector can be found, e.g., in~\cite{2008Yonghong,2009Kritchman,2010Taherpour,2009Lu}. Assuming unknown noise power, the optimal detector in the GLRT sense is the Scaled Largest Eigenvalue based (SLE) detection $\left(T_{\text{SLE}}=\lambda_{1}/\text{tr}(\mathbf{R})\right)$~\cite{2010Wang,2010Bianchi,2011Nadler}.

In the presence of \emph{multiple} primary users ($P\geq 2$), neither LE nor SLE detection are optimal and no uniformly most powerful test exists in this setting~\cite{1982Muirhead}. To formulate a hypothesis test in this setting, one needs to consider the fact that for a secondary user the most critical information is whether or not there are active primary users. The knowledge of the number of active primary users may not be relevant from the secondary user's perspective. With the above considerations and also the fact that $\sum_{i=1}^{P}\gamma_{i}\mathbf{h}_{i}\mathbf{h}^{\dag}_{i}$ is a positive definite matrix, we choose the following hypothesis test in the multiple primary users scenario
\begin{eqnarray}
\mathcal{H}_{0}&:& \mathbf{\Sigma}=\sigma^{2}\mathbf{I}_{K}\\
\mathcal{H}_{1}&:& \mathbf{\Sigma}\succ\sigma^{2}\mathbf{I}_{K},
\end{eqnarray}
where the noise power $\sigma^{2}$ is assumed to be unknown and $\mathbf{A}\succ\mathbf{B}$ denotes that $\mathbf{A}-\mathbf{B}$ is a positive definite matrix.
Essentially, we are now testing a null hypothesis
$\mathbf{\Sigma}=\sigma^{2}\mathbf{I}_{K}$ against all the other
possible alternatives of $\mathbf{\Sigma}$. Thus no a priory
assumption on the structure of $\mathbf{\Sigma}$ is required, except
for its positive definiteness. In
particular, deciding the number of primary users $P$ is not
needed, i.e., the hypothesis test is blind in $P$. Intuitively, this test is to reject
$\mathcal{H}_{0}$ if we have reason to believe that the population covariance matrix $\mathbf{\Sigma}$ departs from the sphericity
$\mathbf{\Sigma}=\sigma^{2}\mathbf{I}_{K}$.

In the statistics literature, this hypothesis test is known as the
sphericity test, which was first studied in~\cite{1940Mauchly}.
Comprehensive results of the sphericity test for the real Wishart matrix $\mathbf{R}$, including asymptotic distributions, can be found in~\cite{1982Muirhead}. The test statistics of this Spherical Test based (ST) detector
was derived under the GLRT criterion as~\cite{1940Mauchly}
\begin{equation}
T_{\text{ST}}=\frac{\det(\mathbf{R})}{\left(\frac{1}{K}\text{tr}(\mathbf{R})\right)^{K}}=\frac{\prod_{i=1}^{K}\lambda_{i}}{\left(\frac{1}{K}\sum_{i=1}^{K}\lambda_{i}\right)^{K}}.
\end{equation}
For completeness, the essential steps of the derivation are outlined
here. Apart from a constant, the likelihood function of the data
matrix $\mathbf{X}$ is
\begin{equation}
L(\mathbf{X}|\mathbf{\Sigma})=\left(\det(\mathbf{\Sigma})\right)^{-N}\e^{\text{tr}(-\mathbf{\Sigma}^{-1}\mathbf{R})}.
\end{equation}
The likelihood ratio statistics is
\begin{equation}\label{eq:GLRT}
\rho:=\frac{\sup_{\sigma^{2}>0}L(\mathbf{X}|\sigma^{2}\mathbf{I}_{K})}{\sup_{\mathbf{\Sigma}\succ0}L(\mathbf{X}|\mathbf{\Sigma})}.
\end{equation}
The maximum likelihood estimates of $\sigma^{2}$ under $\mathcal{H}_{0}$ and
$\mathbf{\Sigma}$ under $\mathcal{H}_{1}$ are~\cite{2003Anderson},
\begin{equation}
\hat{\sigma^{2}}=\frac{\text{tr}(\mathbf{R})}{KN},~~~~\hat{\mathbf{\Sigma}}=\frac{\mathbf{R}}{N}
\end{equation}
respectively. Inserting these into (\ref{eq:GLRT}) we obtain
\begin{equation}
\rho^{1/N}=\frac{\det(\mathbf{R})}{\left(\frac{1}{K}\text{tr}(\mathbf{R})\right)^{K}}:=T_{\text{ST}}.
\end{equation}
Hypothesis $\mathcal{H}_{0}$ is rejected if $\rho$ is small i.e.
when $\rho^{1/N}$ is small. Thus if $T_{\text{ST}}$ is greater than
some threshold $\zeta$, the detector declares $\mathcal{H}_{0}$,
otherwise $\mathcal{H}_{1}$:
\begin{equation}\label{eq:detection}
T_{\text{ST}}\overset{\mathcal{H}_{0}}{\underset{\mathcal{H}_{1}}{\gtrless}}\zeta.
\end{equation}

Recently the spherical test is formulated in~\cite{2010Zhang} as a
spectrum sensing algorithm. However the detection performance
analysis in~\cite{2010Zhang} relies on simulations only. We will address this analytically in the next section. Besides the ST detector, other competing detectors in the presence of multiple primary users include the Eigenvalue Ratio based (ER) detection $\left(T_{\text{ER}}=\lambda_{1}/\lambda_{K}\right)$~\cite{2008Zeng,2009Federico,2010Federico} and John's detection $\left(T_{\text{J}}=\sum_{i=1}^{K}\lambda_{i}^{2}/\left(\sum_{i=1}^{K}\lambda_{i}\right)^{2}\right)$~\cite{1971John,1972John,1972Sugiura}.
In the scenarios simulated in Section~\ref{sec:Simulations} the ST detector outperforms the ER detector in the
presence of multiple primary users and, when $P$ is large, also John's detector.

\section{Performance Analysis}\label{sec:Distributions}

In this section, we derive some closed-form performance metrics for the
spherical test based detection. In the sequel, we present analytical
formulae for its false alarm probability, detection
probability, decision threshold and receiver operating
characteristics.

\subsection{False Alarm Probability}

Define the Cumulative Distribution Function (CDF) of the random
variable $T_{\text{ST}}$ under $\mathcal{H}_{0}$ by
$F_{\text{ST}}(y)$. Since $P_{\text{fa}}$ relies on
$F_{\text{ST}}(y)$, we start by investigating the characteristics of
$F_{\text{ST}}(y)$. For the case of two sensors $K=2$ and three sensors $K=3$ with arbitrary sample size $N$, the exact PDFs of $T_{\text{ST}}$ under
$\mathcal{H}_{0}$ can be found, e.g., in~\cite{1975Nagarsenker} (Eq. (3.8)) and~\cite{1985Nagar} (Eq. (\rmnum2) of Corollary $2.1$) respectively. By definition, the CDFs for $K=2$ and $K=3$ can be obtained as
\begin{equation}\label{eq:K2}
F_{\text{ST}}(y)=\frac{2B_{y}\left(N-1,\frac{3}{2}\right)\Gamma\left(N+\frac{1}{2}\right)}{\sqrt{\pi}\Gamma(N-1)},~~~y\in[0,\infty),
\end{equation}
and
\begin{equation}\label{eq:K3}
F_{\text{ST}}(y)=\frac{\Gamma\left(N+\frac{1}{3}\right)\Gamma\left(N+\frac{2}{3}\right)}{6\Gamma(N-1)\Gamma(N-2)}\sum_{k=0}^{\infty}\frac{\left(\frac{8}{3}\right)_{k}\left(\frac{7}{3}\right)_{k}}{k!(4)_{k}}B_{y}(N-1,k+4),~~~y\in[0,\infty),
\end{equation}
respectively. Here, $B_{y}(\alpha,\beta)=\int_{0}^{y}x^{\alpha-1}(1-x)^{\beta-1}\mathrm{d}x$ is the incomplete Beta function and $\Gamma(\cdot)$ defines the
Gamma function. The Pochhammer symbol $(x)_{n}$ equals $(x)_{n}=\frac{\Gamma(x+n)}{\Gamma(x)}$.

In principle for $K>3$ the exact $T_{\text{ST}}$ distribution can be obtained by the standard approach of Mellin transform~\cite{1982Muirhead}. The resulting density functions may involve the Meijer G-function or the Fox H-function~\cite{1969Consul}. The results, although are of theoretical interest, appear to be of limited usefulness due to their complicated forms.

Since explicit expression for the distribution of $T_{\text{ST}}$ may not be easily obtained for arbitrary $K$, it is more desirable to approximate the distribution by some known distribution, based on fitting the first few moments. In this paper, we choose the Beta distribution since it is defined on the same support $[0,1]$ as the random variable $T_{\text{ST}}$. Additional motivation comes from the fact that the exact density functions in~\cite{1975Nagarsenker} for $K=2$ and in~\cite{1985Nagar} for $K=3$ hold the same polynomial form $x^{i}(1-x)^{j}$ as the Beta density. Accordingly we have

\begin{proposition}\label{th:H0}
For any sensor size $K$ and sample size $N$, the two-first-moment
Beta-approximation to the CDF of $T_{\text{ST}}$ under
$\mathcal{H}_{0}$ is
\begin{equation}\label{eq:CDFH0}
F_{\text{ST}}(y)\approx\frac{B_{y}(\alpha_{0},\beta_{0})}{B(\alpha_{0},\beta_{0})},~~~y\in[0,\infty)
\end{equation}
where
$B(\alpha,\beta)=\frac{\Gamma(\alpha)\Gamma(\beta)}{\Gamma(\alpha+\beta)}$
is the Beta function. The parameters $\alpha_{0}$ and $\beta_{0}$
are given by
\begin{equation}\label{eq:alp0bet0}
\alpha_{0}=\frac{\mathcal{M}_{1}(\mathcal{M}_{1}-\mathcal{M}_{2})}{\mathcal{M}_{2}-\left(\mathcal{M}_{1}\right)^{2}},~~~\beta_{0}=\frac{\left(1-\mathcal{M}_{1}\right)\left(\mathcal{M}_{1}-\mathcal{M}_{2}\right)}{\mathcal{M}_{2}-\left(\mathcal{M}_{1}\right)^{2}}
\end{equation}
with
\begin{equation}
\mathcal{M}_{n}=\frac{\Gamma(KN)}{\Gamma_{K}(N)}\frac{K^{Kn}\Gamma_{K}(N+n)}{\Gamma(K(N+n))}
\end{equation}
where
\begin{equation}\label{eq:multiGam}
\Gamma_{K}(N)=\pi^{\frac{1}{2}K(K-1)}\Gamma(N)\Gamma(N-1)\cdots\Gamma(N-K+1).
\end{equation}
\end{proposition}

The proof of Proposition~\ref{th:H0} is in Appendix~\ref{ap:H0}. Here, $\alpha_{0}$ and $\beta_{0}$ are simple functions of the sensor size $K$ and sample size $N$ only. Note that asymptotic $T_{\text{ST}}$ distributions (w.r.t $N$) for real and complex Wishart matrices can be found in~\cite{1982Muirhead} and~\cite{1990Williams} respectively. Comparison of approximation accuracy of the asymptotic distribution and Proposition~\ref{th:H0} will be performed in Section~\ref{sec:Simulations}.

By~(\ref{eq:detection}), for any threshold $\zeta$ the false alarm
probability is obtained as
\begin{equation}\label{eq:Pfa}
P_{\text{fa}}(\zeta)=F_{\text{ST}}(\zeta).
\end{equation}
Equivalently for any $P_{\text{fa}}$ a threshold can be calculated
by numerically inverting $F_{\text{ST}}(\zeta)$
\begin{equation}\label{eq:thre}
\zeta=F_{\text{ST}}^{-1}(P_{\text{fa}}).
\end{equation}

\subsection{Detection Probability}

Define the CDF of the random variable $T_{\text{ST}}$ under $\mathcal{H}_{1}$ by $G_{\text{ST}}(y)$. Since $P_{\text{d}}$ is related to $G_{\text{ST}}(y)$ we have the following result

\begin{proposition}\label{th:H1K2}
For the case of two sensors $K=2$ with
arbitrary sample size $N$, the exact CDF of $T_{\text{ST}}$ under
$\mathcal{H}_{1}$ is given by
\begin{equation}\label{eq:H1K2}
G_{\text{ST}}(y)=1-C\sum_{k=0}^{\infty}\frac{(3-2N-2k)_{2k-1}}{(2k-1)!}\left(\frac{\sigma_{1}-\sigma_{2}}{\sigma_{1}+\sigma_{2}}\right)^{2k}B_{1-y}(k+\frac{1}{2},N-1),~~~y\in[0,\infty)
\end{equation}
where the constant
\begin{equation}
C=-\frac{4(\sigma_{1}\sigma_{2})^{N}(\sigma_{1}+\sigma_{2})^{2-2N}}{B(N,N-1)(\sigma_{1}-\sigma_{2})^{2}}
\end{equation}
and $\sigma_{i}$ denotes $i$-th eigenvalue of the population
covariance matrix $\mathbf{\Sigma}$~(\ref{eq:coH1}).
\end{proposition}

The proof of Proposition~\ref{th:H1K2} is in Appendix~\ref{ap:H1K2}. Under $\mathcal{H}_{1}$, exact representations for the distribution of $T_{\text{ST}}$ exist in the literature, e.g., Theorem $4.1$ in~\cite{1971Pillai} and Equation $(2.12)$ in~\cite{1971Khatri}. However, whilst being exact for arbitrary $K$ and $N$, these representations involve an infinite sum of products of a Zonal polynomial and the Meijer G-function, which are difficult to compute. The exact distribution of $T_{\text{ST}}$ may not be easily obtained in a computable form when $K>2$. On the other hand, for $K=2$ we see that the distribution is a weighted sum of Beta functions. Since it is a standard technique in statistics to approximate a sum of Betas by one Beta, we extend this to arbitrary $K$ as

\begin{proposition}\label{th:H1}
For any sensor size $K$ and sample size $N$, the two-first-moment
Beta-approximation to the CDF of $T_{\text{ST}}$ under
$\mathcal{H}_{1}$ is
\begin{equation}\label{eq:CDFH1}
G_{\text{ST}}(y)\approx\frac{B_{y}(\alpha_{1},\beta_{1})}{B(\alpha_{1},\beta_{1})},~~~y\in[0,\infty).
\end{equation}
The parameters $\alpha_{1}$ and $\beta_{1}$ are given by
\begin{equation}\label{eq:alp1bet1}
\alpha_{1}=\frac{\mathcal{N}_{1}(\mathcal{N}_{1}-\mathcal{N}_{2})}{\mathcal{N}_{2}-\left(\mathcal{N}_{1}\right)^{2}},~~~~\beta_{1}=\frac{\left(1-\mathcal{N}_{1}\right)\left(\mathcal{N}_{1}-\mathcal{N}_{2}\right)}{\mathcal{N}_{2}-\left(\mathcal{N}_{1}\right)^{2}}
\end{equation}
with
\begin{equation}
\mathcal{N}_{n}=\left(\frac{K}{b}\right)^{Kn}\frac{\Gamma(a-Kn)\Gamma_{K}(N+n)\left(\det(\mathbf{\Sigma})\right)^{n}}{\Gamma_{K}(N)\Gamma(a)}
\end{equation}
where
\begin{equation}\label{eq:GamPara}
a=(N+n)\frac{\left(\sum_{i=1}^{K}\sigma_{i}\right)^{2}}{\sum_{i=1}^{K}\sigma_{i}^{2}},~~~~b=\frac{\sum_{i=1}^{K}\sigma_{i}^{2}}{\sum_{i=1}^{K}\sigma_{i}}.
\end{equation}
\end{proposition}

The proof of Proposition~\ref{th:H1} is in Appendix~\ref{ap:H1}.
Note that results on asymptotic $T_{\text{ST}}$ distribution for real Wishart matrices with arbitrary correlation can be found in~\cite{1982Muirhead}, which may be generalized to the complex Wishart case. However, simulations show that the convergence of these asymptotic distributions can be very slow w.r.t sample sizes for high SNR.

By~(\ref{eq:detection}) the detection probability can be
expressed as
\begin{equation}\label{eq:Pm}
P_{\text{d}}(\zeta)=G_{\text{ST}}(\zeta).
\end{equation}
Note that if we further approximate the parameters
($\alpha_{0}$,$\beta_{0}$) and ($\alpha_{1}$,$\beta_{1}$) to their
respective nearest integer, both (\ref{eq:Pfa}) and (\ref{eq:Pm})
reduce to simple polynomial equations in $\zeta$. Thus the
computational complexity of threshold calculation becomes quite
affordable for on-line implementations.

For a target $P_{\text{fa}}$ we can calculate the resulting
threshold $\zeta$ by~(\ref{eq:thre}). With this threshold the
corresponding $P_{\text{d}}$ can be obtained from~(\ref{eq:Pm}). The
mapping between $P_{\text{fa}}$ and $P_{\text{d}}$ is the so-called
receiver operating characteristics. Thus an analytical ROC
expression for the ST detection can be obtained as
\begin{equation}\label{eq:AppROC}
P_{\text{d}}=G_{\text{ST}}\left(F_{\text{ST}}^{-1}(P_{\text{fa}})\right).
\end{equation}

\subsection{A Note on Approximation Error}

Based on Weierstrass approximation theorem, any square integrable function on a finite interval can be expressed in an orthogonal Jacobi polynomial basis, see e.g.~\cite{1961Hochstadt}. The proposed two-first-moment Beta approximations in Propositions~\ref{th:H0} and~\ref{th:H1} correspond to the simplest form of such an approximation, where two first polynomials matching the moments are used. The approximation error is related to the higher order polynomials left out from the approximation. The functional form of these higher order terms can be found, e.g., in Eq. (6) of~\cite{1993Boik}. In light of Eq. (6) in~\cite{1993Boik}, the exact $P_{\text{fa}}(\zeta)$ and $P_{\text{d}}(\zeta)$ can be written as a sum of the proposed Beta approximation and an error term $e_{\alpha_{i},\beta_{i}}(\zeta)$, which equals
\begin{equation}\label{eq:error}
e_{\alpha_{i},\beta_{i}}(\zeta)=\lim_{M\rightarrow\infty}\left(\sum_{n=3}^{M}A_{n}\sum_{p=1}^{n}\sum_{q=1}^{p}B_{p,q,n}\zeta^{\alpha_{i}+n-q}(1-\zeta)^{\beta_{i}}\right), i=0,1,
\end{equation}
where $A_{n}$ and $B_{p,q,n}$ are some constants. Here, $\alpha_{0}$, $\beta_{0}$ are defined in~(\ref{eq:alp0bet0}) and $\alpha_{1}$, $\beta_{1}$ are defined in~(\ref{eq:alp1bet1}). Due to the complicated form of the error term~(\ref{eq:error}), analysis on its behavior seems difficult. However, in the most interesting cases of low false alarm probability $P_{\text{fa}}(\zeta\rightarrow0)$ and high detection probability $P_{\text{d}}(\zeta\rightarrow1)$, the behavior of the error can be understood. Consider an infinitesimal $\epsilon$ fulfilling $0<\epsilon\ll1$, it follows from~(\ref{eq:error}) that the leading order term in $e_{\alpha_{0},\beta_{0}}(\epsilon)$ for low false alarm probability $P_{\text{fa}}(\epsilon)$ is proportional to $\epsilon^{\alpha_{0}}$ (when $p=q=n$ in~(\ref{eq:error})) and the leading order error in $e_{\alpha_{1},\beta_{1}}(1-\epsilon)$ for high detection probability $P_{\text{d}}(1-\epsilon)$ is $\epsilon^{\beta_{1}}$. Typically, the values $\alpha_{0}$ and $\beta_{1}$ are positive and large. For example, $(\alpha_{0},\beta_{1})$ equals $(395.4,17.0)$, $(195.4,19.1)$ and $(95.5,16.9)$ for the parameters considered in Figure~\ref{fig:ROCP1}, Figure~\ref{fig:ROCP3} and Figure~\ref{fig:ROCP6} respectively. Thus, the corresponding error for low $P_{\text{fa}}$ and high $P_{\text{d}}$ decreases quite fast.

\section{Numerical Results}\label{sec:Simulations}

In this section we first validate the derived approximative
$P_{\text{fa}}$ and $P_{\text{d}}$ expressions by Monte-Carlo
simulations. Then we investigate the performance of ST detection by
comparing with several detection algorithms in the cases with and without noise uncertainty. The considered parameters $K$ and $N$ in this section reflect
practical spectrum sensing scenarios. The sample size $N$ can be as
large as a couple of hundreds whereas the number of sensors $K$ is
typically less than eight due to physical constraints of the device
size. Note that by using the results of~\cite{2009Kritchman,2010Nadler}, a-priori information on the number of primary users may be exploited to improve the detection performance.

\subsection{False Alarm and Detection Probabilities}\label{subsec:P}

\begin{figure}[p]
\centering
\includegraphics[width=3.5in]{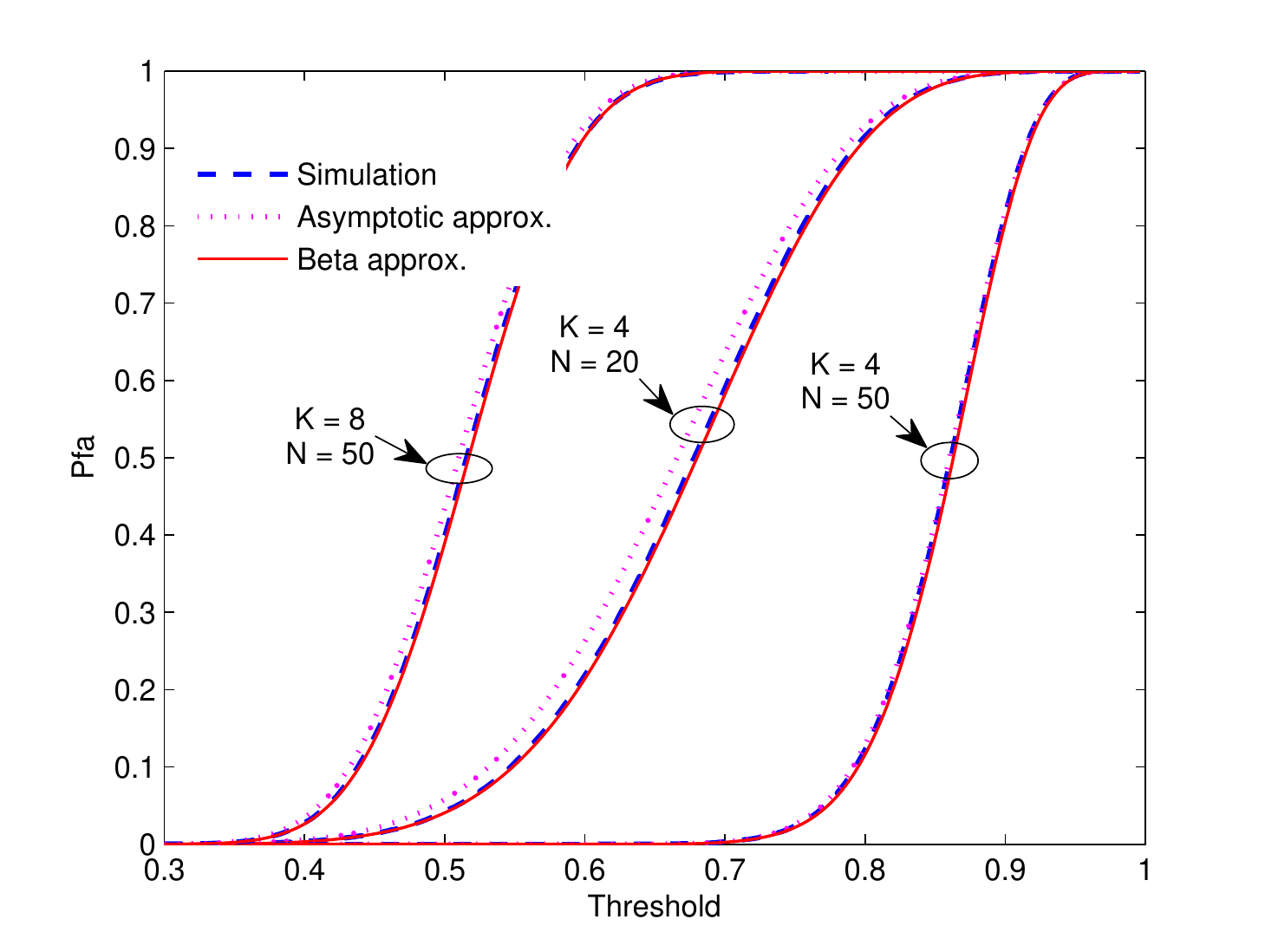}
\caption{False alarm probability: Beta approximation versus
asymptotic distribution~\cite{1990Williams}. For $(K,N)$ values
$(4,20)$, $(4,50)$ and $(8,50)$, the average CDF vertical difference
for the Beta approximation is respectively $4.92\times10^{-8}$,
$5.02\times10^{-8}$, $5.09\times10^{-8}$ and for the asymptotic approximation is respectively $4.14\times10^{-7}$, $4.01\times10^{-8}$, $1.73\times10^{-7}$.}\label{fig:Pfa1}
\end{figure}

In Figure~\ref{fig:Pfa1} we compare the Beta approximated (Proposition~\ref{th:H0}) and the asymptotic~\cite{1990Williams} false alarm probabilities as a function of the threshold for various $K$ and $N$. To quantitatively show the approximation accuracy, we calculate average CDF vertical difference\footnote{For a CDF, $F(x)$, and its estimate $\hat{F}(x)$, the average CDF vertical difference is defined as $\left(\sum_{i=1}^{n}|F(x_{i})-\hat{F}(x_{i})|\right)/n$, where $n$ is the sampling size. Here, we assume uniform sampling in the support of the distribution with $n=10^{7}$.} of the proposed and the asymptotic approximations with respect to the exact distribution as resulting from simulations. The results, summarized in the caption of Figure~\ref{fig:Pfa1}, show that the accuracy of the Beta approximation is not affected much by $K$ and $N$, while the accuracy of the asymptotic distribution increases with $N$ and decreases with $K$, as expected. For $(K,N) = (4,20)$, the Beta approximation is  an order of magnitude better than the asymptotic result.

\begin{figure}[p]
\centering
\includegraphics[width=3.5in]{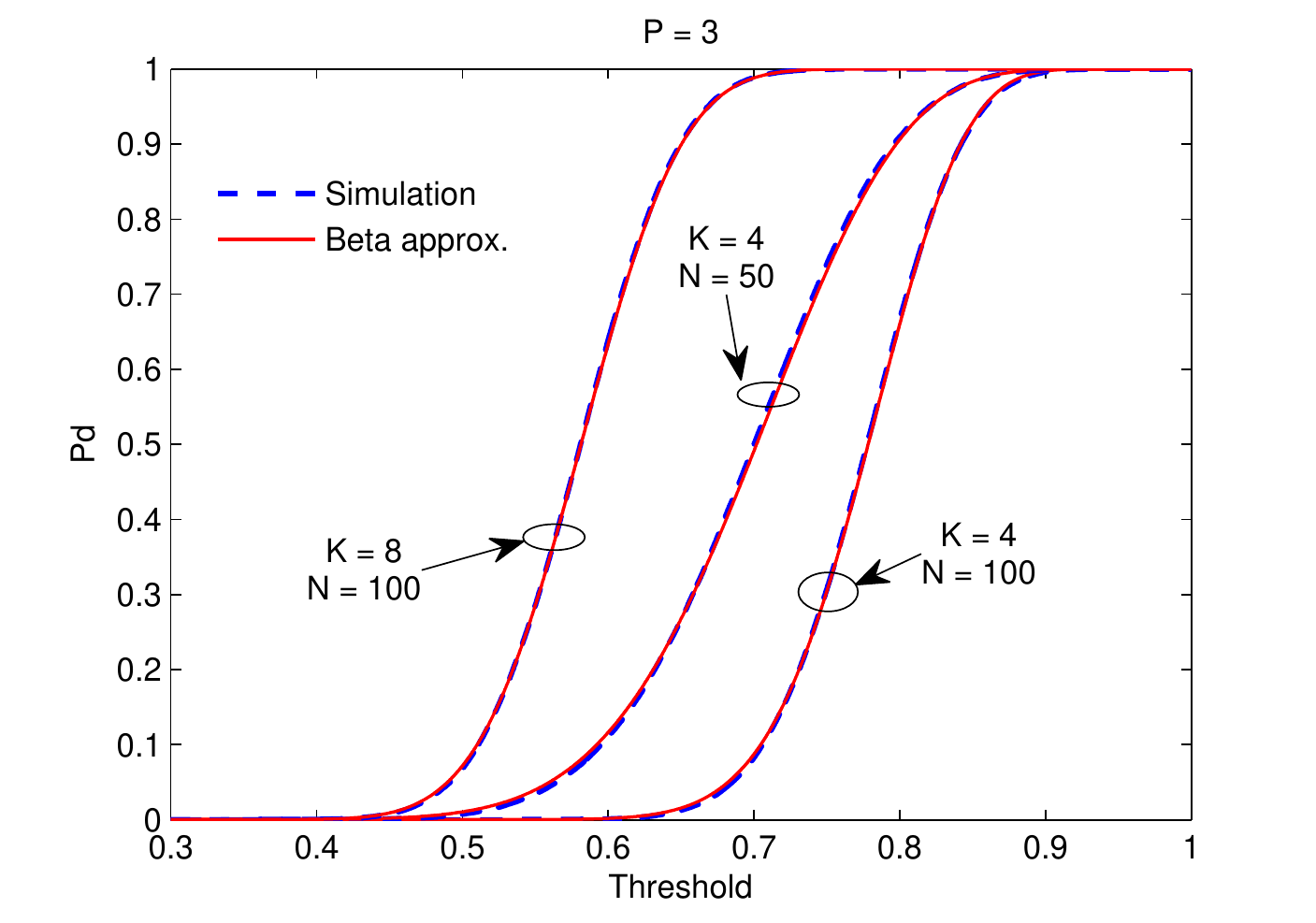}
\caption{Detection probability (assuming three
primary users with $\text{SNR}_{1}=-1$~dB, $\text{SNR}_{2}=-3$~dB
and $\text{SNR}_{3}=-10$~dB): Beta approximation versus simulation.}\label{fig:Pd2}
\end{figure}

In Figure~\ref{fig:Pd2} we plot the derived analytical detection probability versus simulations, assuming three simultaneously transmitting primary users ($P=3$) with $\text{SNR}_{1}=-1$~dB, $\text{SNR}_{2}=-3$~dB and $\text{SNR}_{3}=-10$~dB. Our focus here is detection in the low SNR regime, which is a practical and challenging issue in cooperative spectrum sensing. The Beta approximated $P_{\text{d}}$ curves are calculated using~(\ref{eq:Pm}), where the entries of the channel matrix $\mathbf{H}$ are independently drawn from a standard complex Gaussian distribution corresponding to Rayleigh fading. The channel is fixed during sensing and is normalized as $\mathbf{u}_{i}=\mathbf{h}_{i}/||\mathbf{h}_{i}||$. Without loss of generality, we set the powers of the zero mean Gaussian signal and noise to be $1$. Thus, the population covariance matrix $\mathbf{\Sigma}$ can be explicitly represented as a function of SNRs, i.e, $\mathbf{\Sigma}=\mathbf{I}_{K}+\sum_{i=1}^{P}\text{SNR}_{i}\mathbf{u}_{i}\mathbf{u}^{\dag}_{i}$. With the same $\mathbf{\Sigma}$, the corresponding simulated curve is plotted using $10^{5}$ Monte Carlo runs. From Figure~\ref{fig:Pd2} it can be observed that the derived $P_{\text{d}}$ expression agrees with the simulations well.

\subsection{Detection Performance}

\begin{figure*}[t!]
\centering
\subfigure[$P=1$; $\mu=0$~dB]{\includegraphics[width=2.35in]{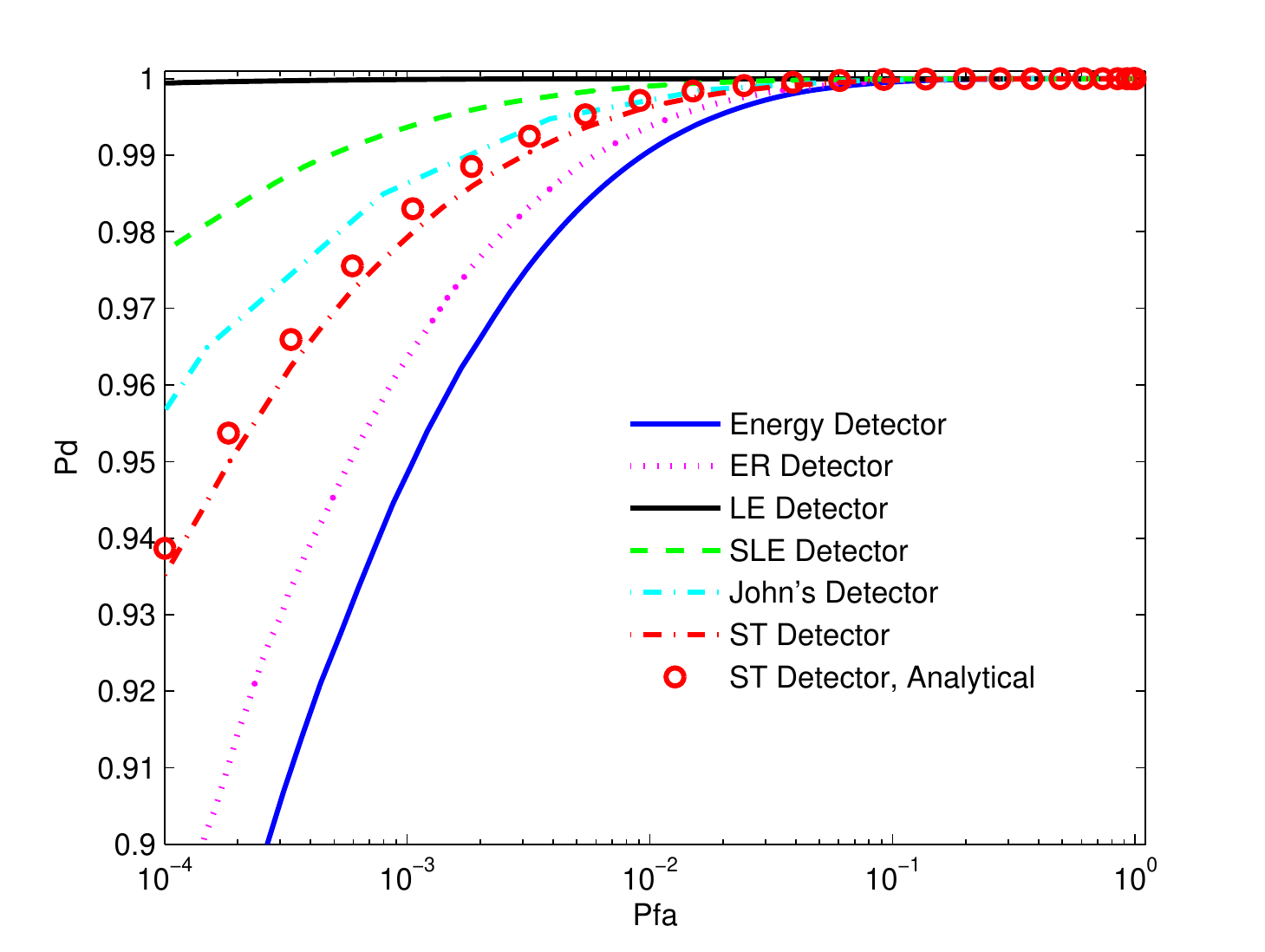}}\subfigure[$P=1$; $\mu=0.5$~dB]{\includegraphics[width=2.35in]{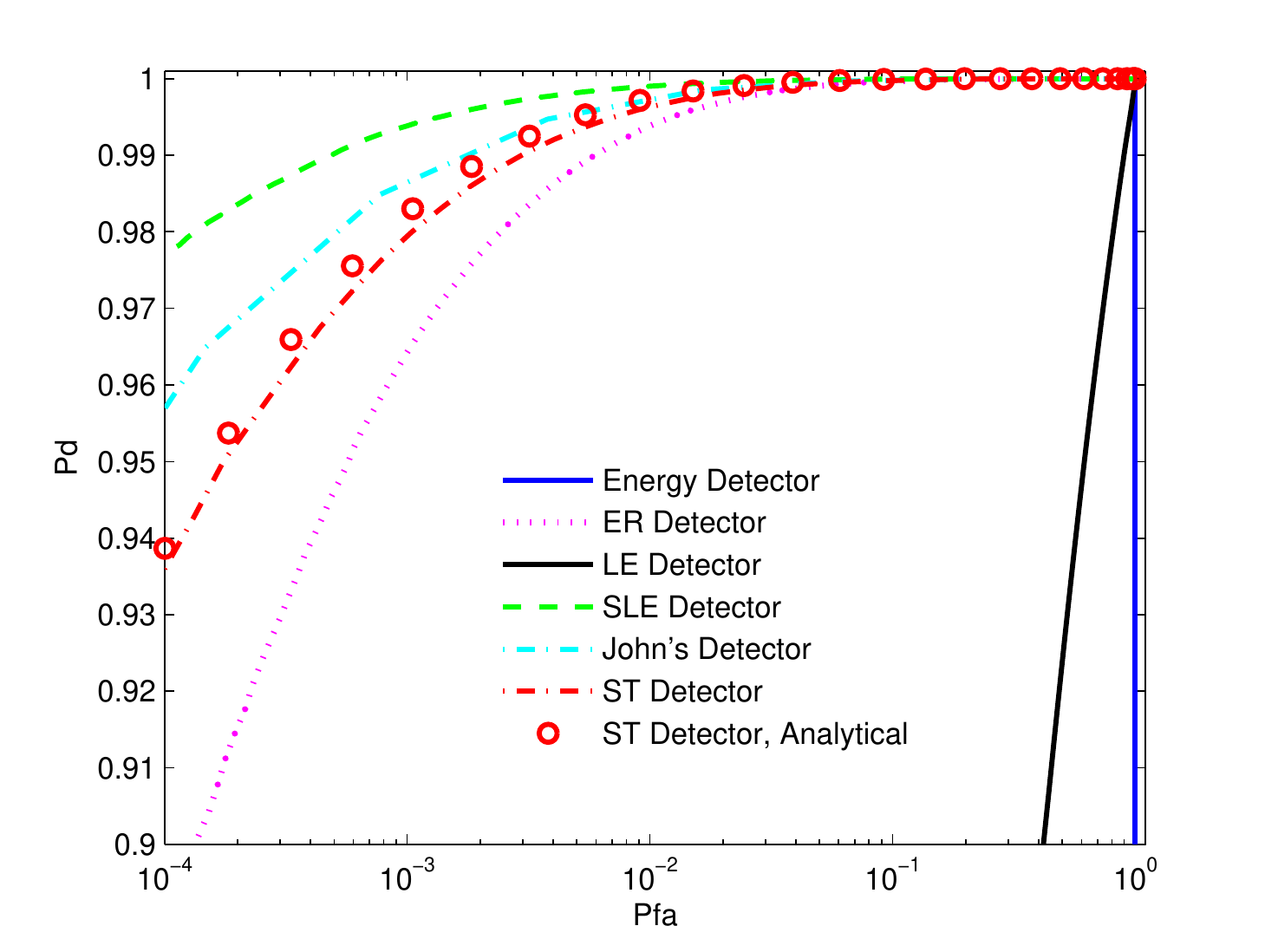}}\subfigure[$P=1$; $\mu=1$~dB]{\includegraphics[width=2.35in]{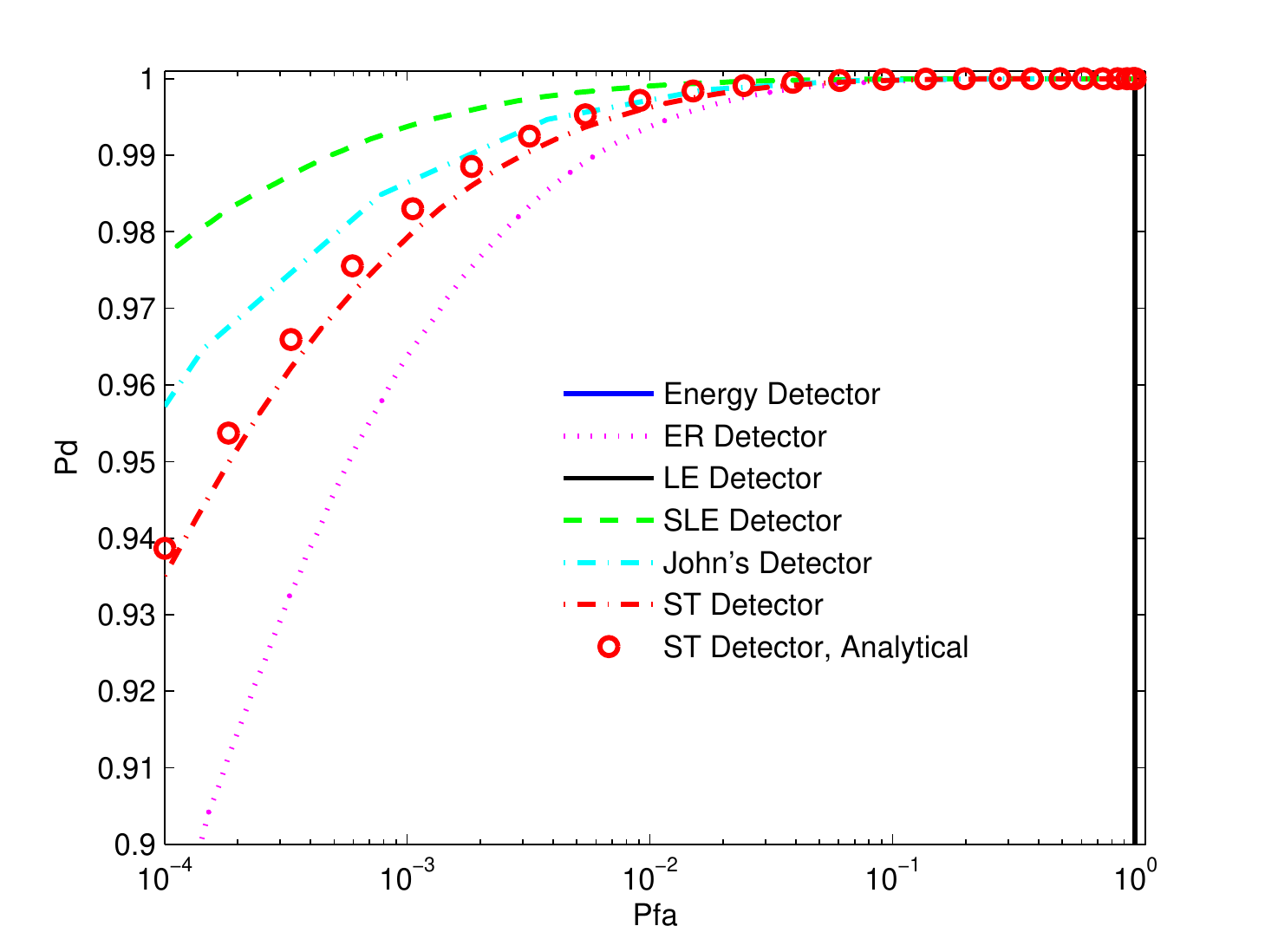}}
\caption{ROC: assuming one primary user with $\text{SNR}_{1}=-3$~dB. The parameters are $K=4$ and $N=400$.}
\label{fig:ROCP1}
\end{figure*}

\begin{figure*}[t!]
\centering
\subfigure[$P=3$; $\mu=0$~dB]{\includegraphics[width=2.35in]{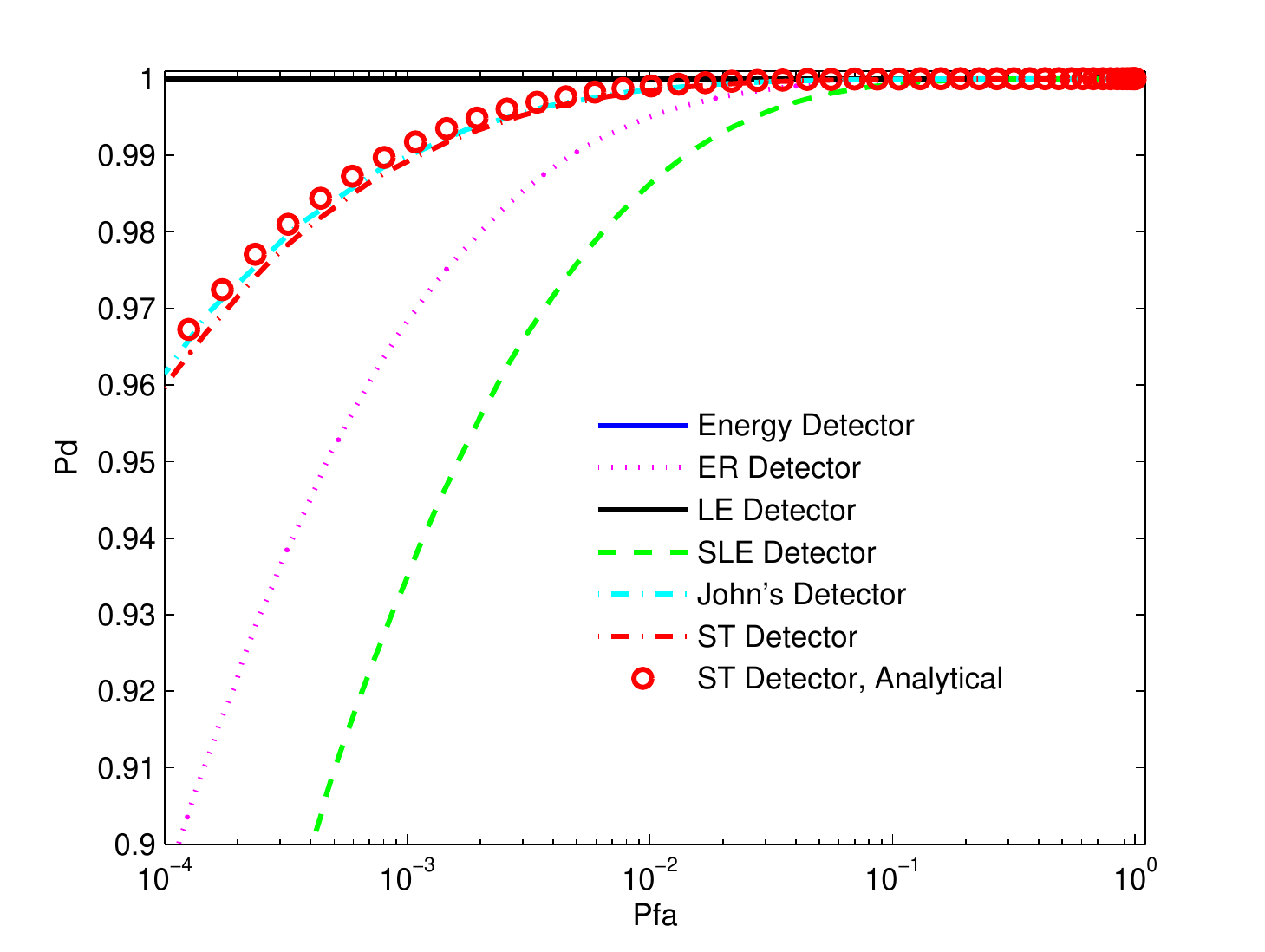}}\subfigure[$P=3$; $\mu=0.5$~dB]{\includegraphics[width=2.35in]{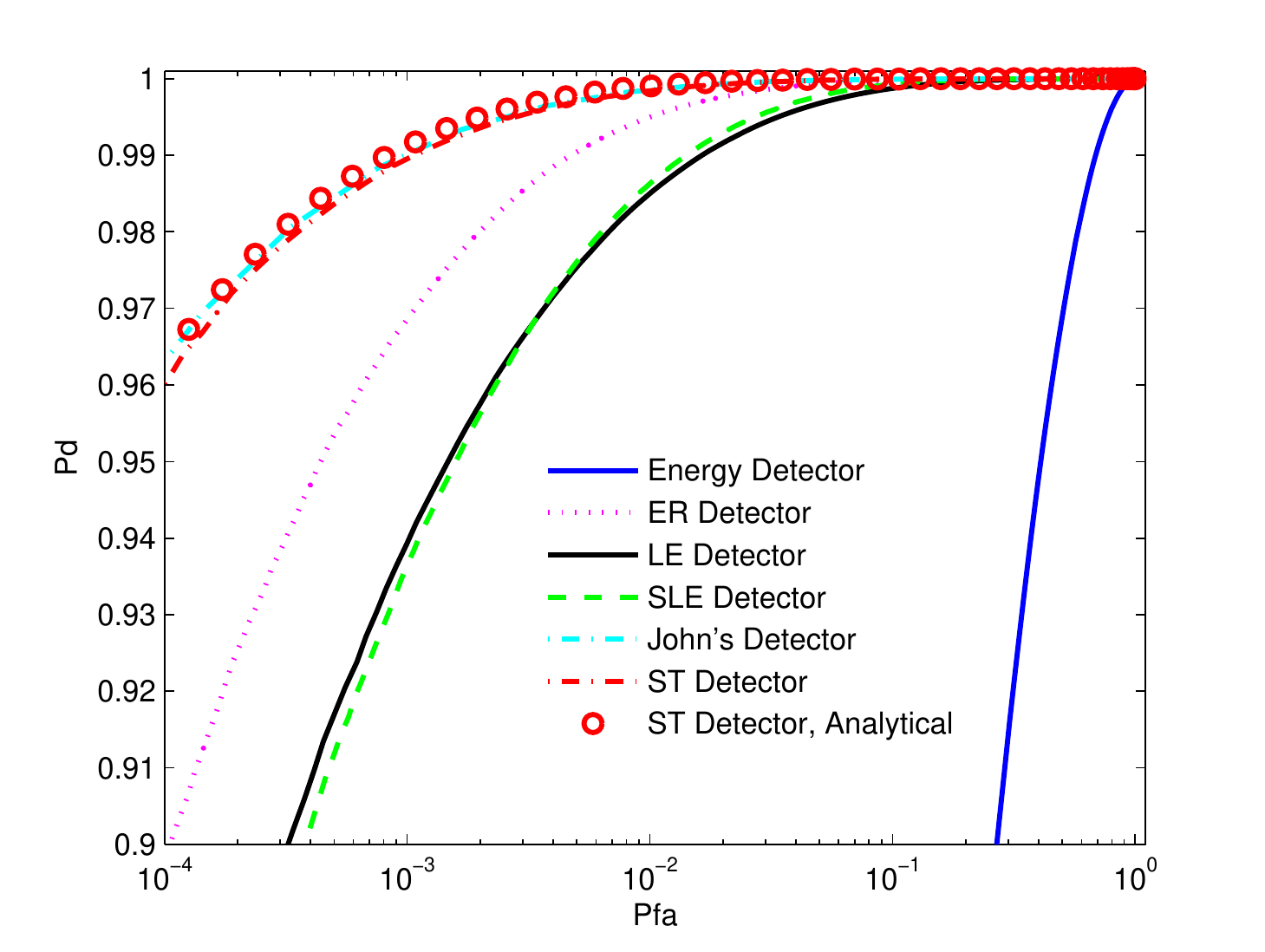}}\subfigure[$P=3$; $\mu=1$~dB]{\includegraphics[width=2.35in]{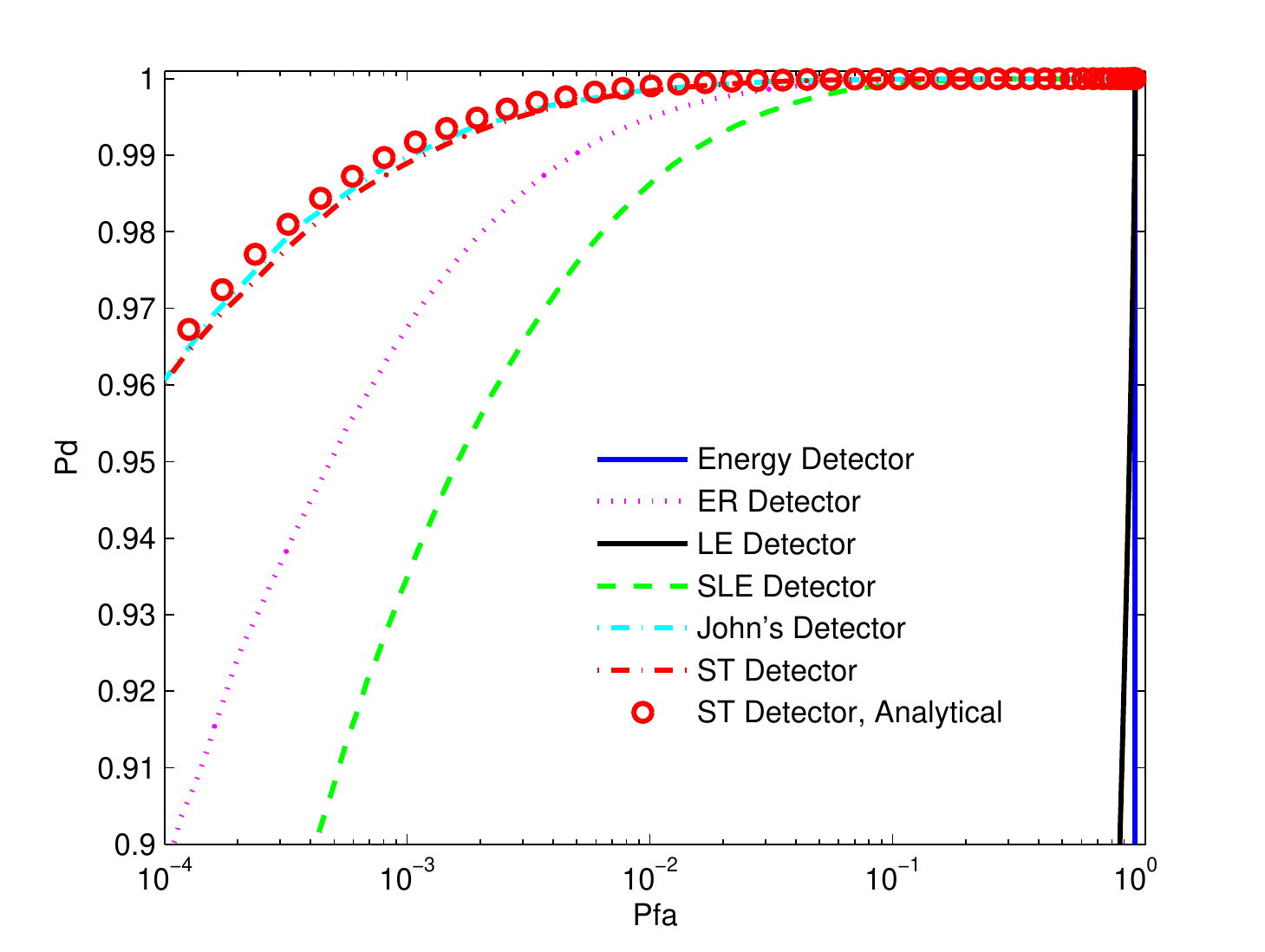}}
\caption{ROC: assuming three primary users with $\text{SNR}_{1}=-1$~dB, $\text{SNR}_{2}=-3$~dB and $\text{SNR}_{3}=-10$~dB. The parameters are $K=4$ and $N=200$.}
\label{fig:ROCP3}
\end{figure*}

\begin{figure*}[t!]
\centering
\subfigure[$P=6$; $\mu=0$~dB]{\includegraphics[width=2.35in]{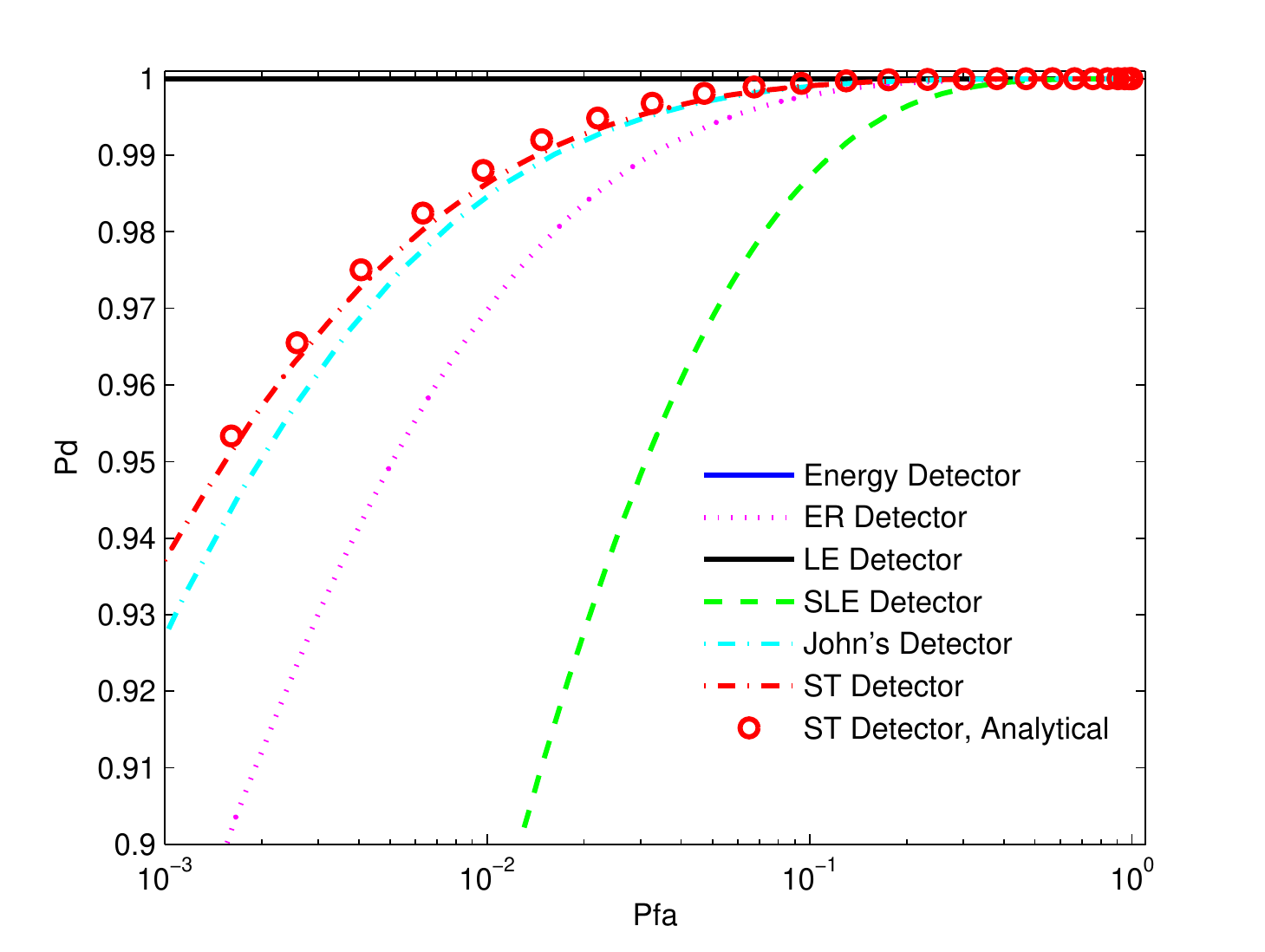}}\subfigure[$P=6$; $\mu=0.5$~dB]{\includegraphics[width=2.35in]{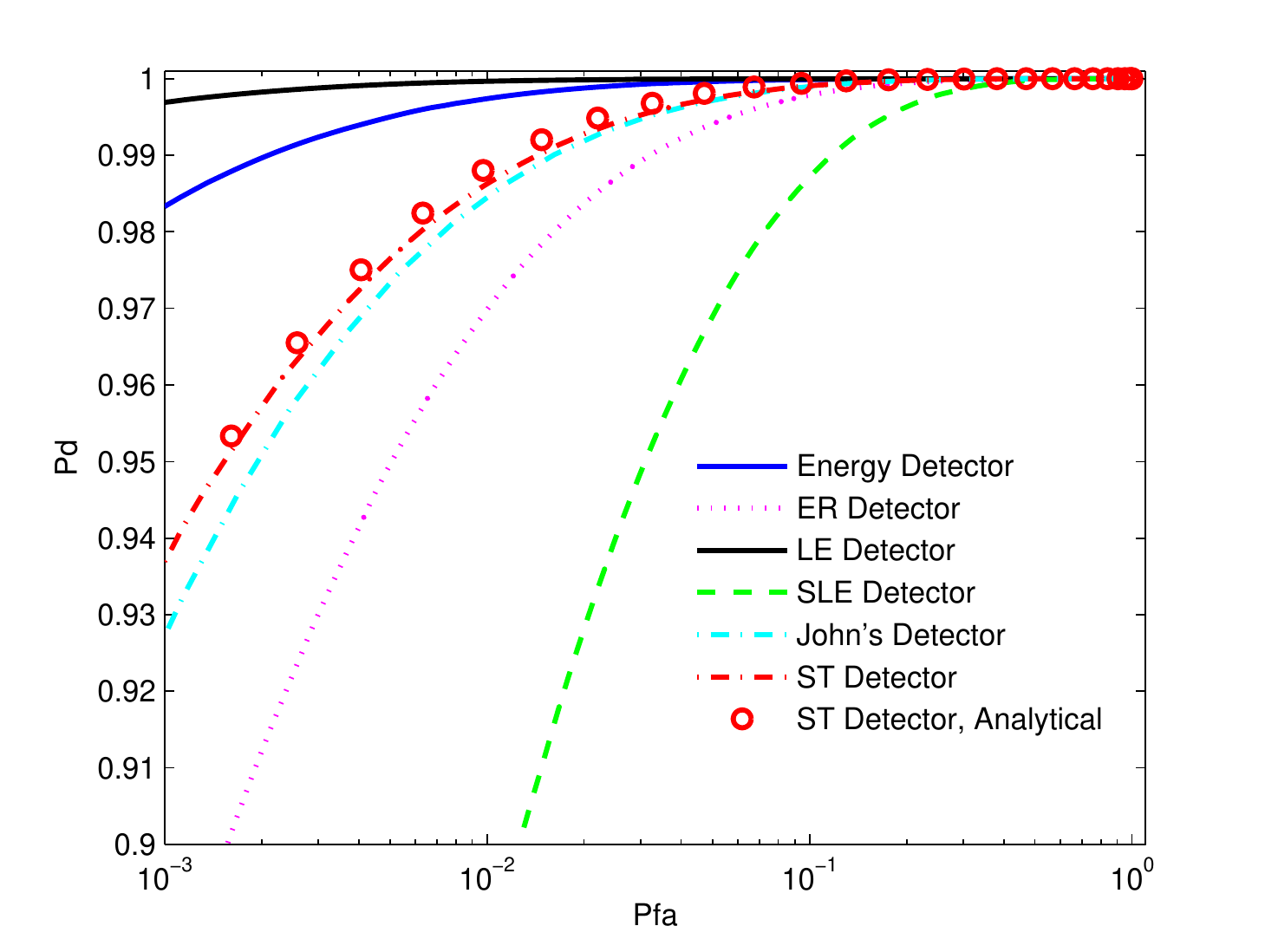}}\subfigure[$P=6$; $\mu=1$~dB]{\includegraphics[width=2.35in]{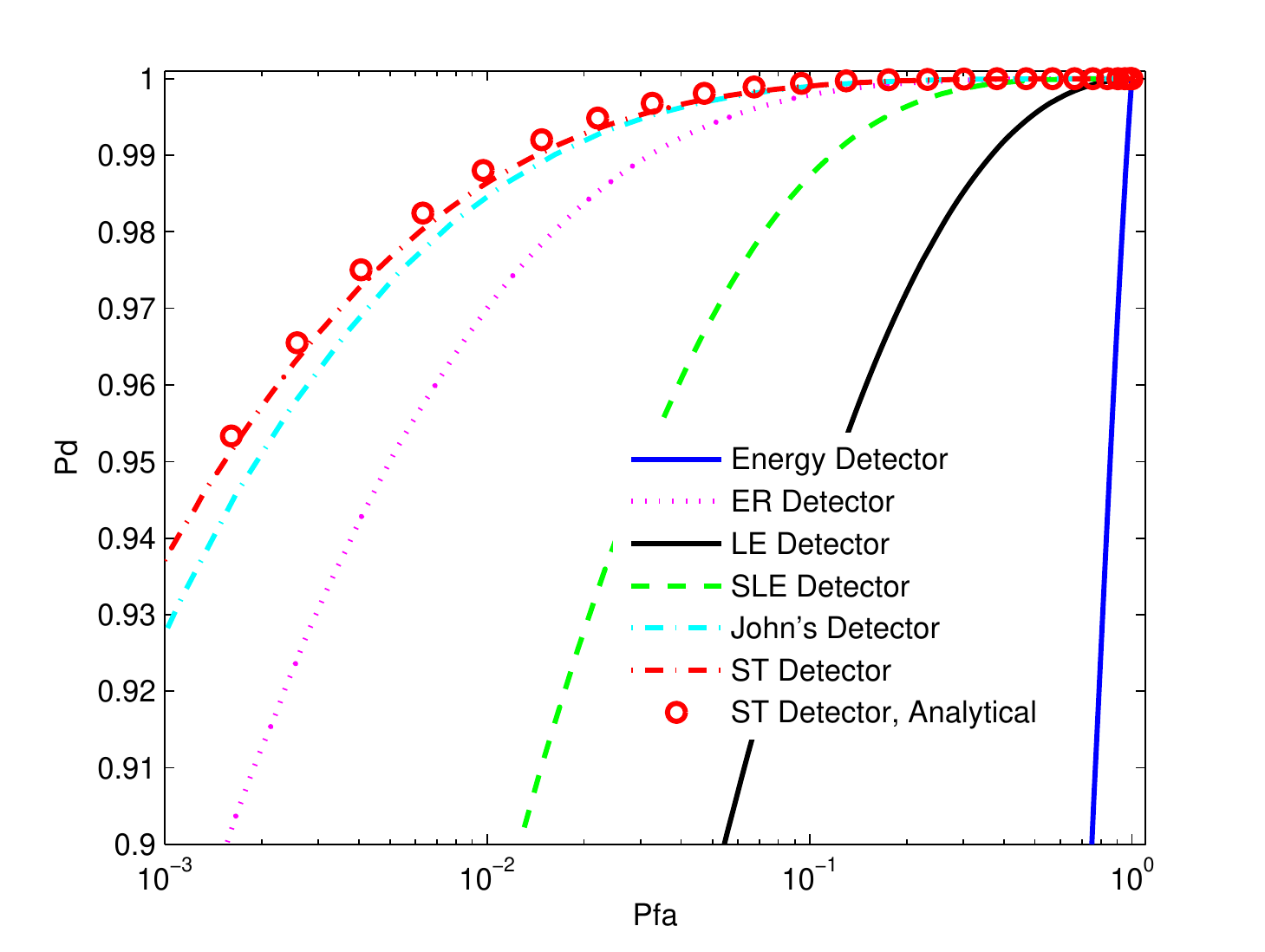}}
\caption{ROC: assuming six primary users with $\text{SNR}_{1}=0$~dB, $\text{SNR}_{2}=-1$~dB, $\text{SNR}_{3}=-3$~dB, $\text{SNR}_{4}=-8$~dB, $\text{SNR}_{5}=-10$~dB and $\text{SNR}_{6}=-22$~dB. The parameters are $K=4$ and $N=100$.}
\label{fig:ROCP6}
\end{figure*}

We compare the detection performance of the spherical test based
detector with other known detectors by means of the ROC curves.
Since the ROC curve shows the achieved detection probability
as a function of the false alarm probability, it reflects the
overall detection performance for a given detector. We consider for
comparison the cooperative Energy Detector
$T_{\text{ED}}=||\mathbf{X}||^{2}_{F}$\footnote{$||\mathbf{\cdot}||_{F}$
denotes the Frobenius norm.}~\cite{2003Digham,2005Sahai}. In addition, the previously discussed Eigenvalue Ratio based detector $T_{\text{ER}}=\lambda_{1}/\lambda_{K}$ and John's detector $T_{\text{J}}=\sum_{i=1}^{K}\lambda_{i}^{2}/\left(\sum_{i=1}^{K}\lambda_{i}\right)^{2}$, which are candidate detectors in the presence of multiple primary users, are compared to. Optimal detectors for a single primary user, such as the LE detector $T_{\text{LE}}=\lambda_{1}$ and the SLE detector $T_{\text{SLE}}=\lambda_{1}/\text{tr}(\mathbf{R})$ are considered for comparison as well.

In practical systems we may not have perfect knowledge of the noise
power~\cite{2005Sahai,2008Tandra}. The noise uncertainty may arise due to interference, noise estimation errors or non-linearity of components~\cite{2005Sahai}. Hence for practical systems, modeling the noise uncertainty is unavoidable. The energy detector and the LE detector are subject to noise uncertainty due to dependence of the test statistics on the noise power~\cite{2005Sahai,2011Nadler}. The SLE, ER, John's and the ST detectors are immune to noise uncertainty since the noise powers are replaced by their respective ML estimates in
constructing the test statistics. Robustness to noise uncertainty is
of fundamental importance because the uncertainty may severely
degrade detection performance, especially at low SNR. If $\mu$
denotes the degree of noise uncertainty in dB, the actual noise
power thus falls in the interval
$\Omega=[\sigma^{2}/\rho,\rho\sigma^{2}]$, where $\rho=10^{\mu/10}$.
As the same in~\cite{2005Sahai,2008Tandra,1992Sonnenschein},
in the following plots we consider the worst performance degradation
due to noise uncertainty, where the noise power is $\rho\sigma^{2}$
under $\mathcal{H}_{0}$ and $\sigma^{2}/\rho$ under
$\mathcal{H}_{1}$. This noise uncertainty model is not only of theoretical interest~\cite{1992Sonnenschein} but also realistic~\cite{2005Sahai}. For example, in order to protect the primary system and guarantee the quality of service for the secondary system, design margins on $P_{\text{fa}}$ and $P_{\text{m}}$ shall be imposed which can be only obtained from the worst case noise uncertainty analysis~\cite{1992Sonnenschein}. Note that the considered noise uncertainty levels here, $0.5$~dB and $1$~dB, are generally realistic in spectrum sensing scenarios. For example, it was remarked in~\cite{2005Sahai} that the noise uncertainty can be at least $1$-$2$~dB due to limitations of devices only and in~\cite{{1992Sonnenschein}} the authors considered noise uncertainty levels up to $3$~dB.

In order to see a clear picture of the impact of number of primary users $P$ and the noise uncertainty $\mu$ on the detection performance, we plot various ROC curves assuming different $P$ and $\mu$. In Figure~\ref{fig:ROCP1} we show the performance of various detectors in the presence of a single primary user with $\text{SNR}_{1}=-3$~dB using $K=4$ sensors and $N=400$ samples per sensor. In Figure~\ref{fig:ROCP3} we assume a scenario of three simultaneously transmitting primary users with $\text{SNR}_{1}=-1$~dB, $\text{SNR}_{2}=-3$~dB and $\text{SNR}_{3}=-10$~dB. The number of sensors is chosen to be $K=4$, and $N=200$ samples per sensor are considered. The ST detector works also when the number of active primary users $P$ is larger than the number of sensors $K$. This fact is illustrated in Figure~\ref{fig:ROCP3}, where we assume the existence of six primary users with $\text{SNR}_{1}=0$~dB, $\text{SNR}_{2}=-1$~dB, $\text{SNR}_{3}=-3$~dB, $\text{SNR}_{4}=-8$~dB, $\text{SNR}_{5}=-10$~dB and $\text{SNR}_{6}=-22$~dB. We consider four cooperating sensors $K=4$ with sample size $N=100$. The analytical approximation of the ROC curves are obtained by~(\ref{eq:AppROC}), where we assume Rayleigh fading channels which are kept constant during sensing. With the same channel realizations, the corresponding numerical ROC is plotted as follows. Without loss of generality, we set the noise power $\sigma^{2}=1$ at the secondary receiver. At noise uncertainty level $\mu=0$~dB (no uncertainty), $\mu=0.5$~dB and $\mu=1$~dB, the noise power becomes $1$ (no uncertainty), $1.122$ and $1.259$ respectively under $\mathcal{H}_{0}$ and $1$ (no uncertainty), $0.891$ and $0.794$ respectively under $\mathcal{H}_{1}$. For each ROC curve, $5\times10^{6}$ realizations of the data matrix $\mathbf{X}$ are drawn from a standard uncorrelated and correlated complex Gaussian distribution under $\mathcal{H}_{0}$ and $\mathcal{H}_{1}$ respectively with the corresponding noise powers given above. For each realization, the test statistics for the considered detectors are calculated under both hypotheses, from which the empirical test statistics distributions are obtained. Using the empirical distributions, the simulated ROC curves are constructed by comparing with $1000$ equally spaced thresholds in the domain of each test statistics.

\emph{Remarks on comparisons with John's detector}:
We observe in Figure~\ref{fig:ROCP1} that John's detector outperforms the ST detector in the presence of a single primary user ($P=1$). In this case, $\mathbf{\Sigma}$ equals an identity matrix plus a rank one perturbation, where the strength of this perturbation is specified by the SNR. With an increased number of primary users ($P=3$), the eigenvalues of $\mathbf{\Sigma}$ become more distinct from each other, where it can be seen from Figure~\ref{fig:ROCP3} that the ST and John's detectors perform almost equally well. When the number of primary users further increases ($P=6$), the eigenvalues of $\mathbf{\Sigma}$ become even more spread. In this case we see from Figure~\ref{fig:ROCP6} that the ST detector outperforms John's detector. To further investigate their relative performance, we simulated their detection probabilities as a function of SNR, where the false alarm probability is set at $10^{-2}$. We assume the presence of two active primary users ($P=2$) with the difference of their SNR fixed: $\text{SNR}_{2}=\text{SNR}_{1}-2$~dB using $K=4$ sensors and $N=50$ samples per sensor. The result is summarized in Table~\ref{table:PdSNR}, where blue color indicates the higher $P_{\text{d}}$ for a given SNR. From Table~\ref{table:PdSNR} we observe that when SNRs of the primary users increase (eigenvalues of $\mathbf{\Sigma}$ become more distinct), ST detector achieves better performance than John's detector, though the differences are small.

The above observations are consistent with the those in~\cite{1984Grieve,1988Chan,1990Boik}, where performance comparisons of John's and ST detectors were made. A common conclusion in~\cite{1984Grieve,1988Chan,1990Boik} is that the relative performance of John's and ST detectors depends on the rank of the perturbation matrix\footnote{The perturbation matrix refers to the difference of the population covariance matrices under $\mathcal{H}_{1}$ and $\mathcal{H}_{0}$, which, in our case, equals $\sum_{i=1}^{P}\gamma_{i}\mathbf{h}_{i}\mathbf{h}^{\dag}_{i}$.} (in our setting the value of $P$) and the distinctness of its eigenvalues (in our setting it depends on the SNRs). A complete understanding of the conditions under which John's detector outperforms the ST detector, or vice versa, seems difficult partially due to the absence of a computable and accurate ROC for John's test. Despite this, more detailed understanding and consequently some general recommendations on the use of the tests can be made based on results of~\cite{1984Grieve,1988Chan,1990Boik}, and our observations.
\begin{itemize}
  \item In the case of two sensors, John's and ST detectors achieve the same performance since their test statistics, up to a linear transform, are the same when $K=2$~\cite{1984Grieve}.
  \item The performance gap of these detectors is not expected to be large~\cite{1984Grieve,1988Chan,1990Boik}, and their asymptotic performance is the same (measured by the Pitman efficiency)~\cite{1984Grieve}.
  \item With one active primary user (rank $1$ perturbation matrix), John's detector is preferable~\cite{1984Grieve,1988Chan,1990Boik}.
  \item When the number of primary users is large (compared to the number of sensors) with not-too-low SNRs (distinct eigenvalues of $\mathbf{\Sigma}$), the ST detector is preferable~\cite{1984Grieve,1988Chan,1990Boik}.
\end{itemize}
John's test is the so-called locally best invariant test, i.e., it is the most powerful test in the neighborhood of $\mathcal{H}_{0}$, although the neighborhood in which it is best is small~\cite{1990Boik}. In our setting, this effectively requires that the sum of SNRs is small~\cite{1990Boik} and the number of primary users is not too large (compared to sensor sizes)~\cite{1988Chan}. Since low SNR detection is of great interest, John's test is a viable alternative in the multiple primary users setting, despite that~\cite{1988Chan} concluded by recommending the use of the ST detector in general.

\begin{table*}[t!]
\caption{Detection probability as a function of SNR} \centering
\begin{tabular}{|c|c|c|c|c|c|c|c|c|c|}
\hline
$\text{SNR}_{1}$ in dB & -1 & -0.5 & 0 & 0.5 & 1 & 1.5 & 2 & 2.5 & 3 \\

\hline
\hline

$P_{\text{d}}$ of ST detector & $0.3628$ & $0.4668$ & $0.5891$ & \textcolor{blue}{$0.7119$} & \textcolor{blue}{$0.8105$} & \textcolor{blue}{$0.8939$} & \textcolor{blue}{$0.9482$} & \textcolor{blue}{$0.9817$} & \textcolor{blue}{$0.9935$} \\

\hline

$P_{\text{d}}$ of John's detector & \textcolor{blue}{$0.3721$} & \textcolor{blue}{$0.4745$} & \textcolor{blue}{$0.5901$} & $0.7057$ & $0.8094$ & $0.8910$ & $0.9458$ & $0.9781$ & $0.9929$ \\

\hline
\end{tabular}
\label{table:PdSNR}
\end{table*}

\emph{Remarks on comparisons with the SLE and the ER detectors}:
In Figure~\ref{fig:ROCP1} we observe that the SLE detector has the best detection performance among the noise uncertainty free detectors considered. Indeed the SLE detector is proved to be optimal for single primary detection under the GLRT criterion~\cite{2010Wang}. When the number of the active primary users is more than one, we see from Figure~\ref{fig:ROCP3} and Figure~\ref{fig:ROCP6} that the ST detector performs better than the SLE detector. It can be observed that the ST detector always achieves better performance than the ER detector. This is intuitively clear by examining their test statistics. For the ER detector, the test statistics depends only on the extreme eigenvalues of the sample covariance matrix $\mathbf{R}$, whereas the test statistics of the ST detector is a function of all the eigenvalues of $\mathbf{R}$.

\emph{Remarks on comparisons with the LE and the ED detectors}:
It can be observed from subplot (a) of Figure~\ref{fig:ROCP1},~\ref{fig:ROCP3} and~\ref{fig:ROCP6} that, in cases of perfectly estimated noise power, the ED and LE detectors almost always outperform the ST detector. However, the performance of ED and LE detectors are very sensitive to noise uncertainty which is particularly true when the number of the primary users are small. For example, from Figure~\ref{fig:ROCP1} (b) and Figure~\ref{fig:ROCP3} (c) we can see that both ED and LE detectors fail at $\mu=0.5$~dB when $P=1$ and at $\mu=1$~dB when $P=3$. Note that in practice noise uncertainty is always present~\cite{2005Sahai} thus the superb performance for the ED and LE detectors in the subplot (a) of each figure may not be achieved in real world scenarios.

\section{Conclusion and Future Work}\label{sec:Conclusion}

In this paper, we investigated the sensing performance of a multiple
primary users detector, based on the spherical test. The ST detector
estimates whether the covariance matrix differs from a matrix
proportional to identity. Analytical formulae have been found for
the key performance metrics of the ST detector. For generic values
of the number of sensors $K$, the formulae are based on a
two-first-moment Beta-approximation to the corresponding CDFs. The
derived results are simple to calculate and yield an almost exact
fit to simulations. From the simulation setting considered, performance gain over several detection algorithms is observed in scenarios with noise uncertainty and large number of primary users.

The key message of this paper is that in the presence of more than one primary users, some performance gain may be obtained via the spherical test even without knowing the number of primary users. With only one primary user, however, the LE and SLE detectors prove to be optimal under the GLRT criterion when the noise power is known and unknown respectively. Naturally, one can argue that if some a-priori information on whether $P=1$ or $P>1$ is available, then by switching between the ST and the SLE (or the LE if the noise power is known) detection algorithms, advantages of these detectors can be dynamically exploited. The a-priori information may be acquired by utilizing the recent advances in estimation algorithm for $P$~\cite{2009Kritchman,2010Nadler}. How to analytically capture the above `estimation-assisted detector' remains as interesting future work.

\section*{Acknowledgment}

This first author wishes to thank Prof. Dietrich von Rosen and Dr. Boaz Nadler for their enlightening discussions on this topic. The authors wish to thank the reviewers and the editor for the constructive comments which significantly improve this paper.

\appendices

\section{Distribution of $T_{\text{ST}}$ Under
$\mathcal{H}_{0}$}\label{ap:H0}

Here we prove Proposition~\ref{th:H0}. We first derive the exact moments of
$T_{\text{ST}}$, which is valid for any $K$ and $N$. Define the
random variable $T_{\text{ST}}$ by
\begin{equation}
X:=\frac{\det(\mathbf{R})}{\left(\frac{1}{K}\text{tr}(\mathbf{R})\right)^{K}},
\end{equation}
where it can be verified that $x\in[0,1]$. Under $\mathcal{H}_{0}$,
the sample covariance matrix $\mathbf{R}$ follows an uncorrelated
complex Wishart distribution
$\mathcal{W}_{K}\left(N,\mathbf{I}_{K}\right)$ with density
function\footnote{Since $T_{\text{ST}}$ is independent of
$\sigma^{2}$, without loss of generality we set $\sigma^{2}=1$.}
\begin{equation}
\mathbf{R}\sim
\frac{1}{\Gamma_{K}(N)}\left(\det(\mathbf{R})\right)^{N-K}\e^{\text{tr}(-\mathbf{R})},
\end{equation}
where $\Gamma_{K}(N)$ is defined in~(\ref{eq:multiGam}). Since $X$
is a scalar function of matrix argument $\mathbf{R}$, its $n$-th
moment can be calculated as
\begin{eqnarray}
\mathbb{E}[x^{n}]&=&\frac{K^{Kn}}{\Gamma_{K}(N)}\int_{\mathbf{R}\succ0}\left(\det(\mathbf{R})\right)^{N-K+n}\e^{\text{tr}(-\mathbf{R})}\left(\text{tr}(\mathbf{R})\right)^{-Kn}\mathrm{d}\mathbf{R}\\
&=&\frac{K^{Kn}\Gamma_{K}(N+n)}{\Gamma_{K}(N)}\int_{\mathbf{R}\succ0}\frac{\left(\det(\mathbf{R})\right)^{N-K+n}\e^{\text{tr}(-\mathbf{R})}}{\Gamma_{K}(N+n)}\left(\text{tr}(\mathbf{R})\right)^{-Kn}\mathrm{d}\mathbf{R}\\
&=&\frac{K^{Kn}\Gamma_{K}(N+n)}{\Gamma_{K}(N)}\mathbb{E}[\left(\text{tr}(\mathbf{R'})\right)^{-Kn}],
\end{eqnarray}
where the last expectation is with respect to the Wishart matrix
$\mathbf{R'}$ distributed as
$\mathcal{W}_{K}\left(N+n,\mathbf{I}_{K}\right)$. The random variable
$2\text{tr}(\mathbf{R'})$ follows a Chi-square distribution with $2K(N+n)$ degrees
of freedom, by using the moment expression for Chi-square distribution~\cite{2002Simon} (Eq. (2.35)), the $(-Kn)$-th moment of $\text{tr}(\mathbf{R'})$ is obtained as
\begin{equation}
\mathbb{E}[\left(\text{tr}(\mathbf{R'})\right)^{-Kn}]=\frac{\Gamma(KN)}{\Gamma\left(K(N+n)\right)}.
\end{equation}
The $n$-th moment of $X$ is now
\begin{equation}\label{eq:MoH0}
\mathbb{E}[x^{n}]=\frac{\Gamma(KN)}{\Gamma_{K}(N)}\frac{K^{Kn}\Gamma_{K}(N+n)}{\Gamma(K(N+n))}:=\mathcal{M}_{n}.
\end{equation}
Note that the expression for the exact moments can be also obtained
by exploiting the independence between random variables $X$ and
$\text{tr}(\mathbf{R})$ under $\mathcal{H}_{0}$~\cite{1940Mauchly}.

The first two moments of
$T_{\text{ST}}$ can be obtained by using~(\ref{eq:MoH0}). For a
Beta distribution with density function
\begin{equation}
\frac{1}{B(\alpha_{0},\beta_{0})}x^{\alpha_{0}-1}(1-x)^{\beta_{0}-1},~~~x\in[0,1]
\end{equation}
equaling the first two moments to the moments of $T_{\text{ST}}$ we
have
\begin{equation}
\mathcal{M}_{1}=\frac{\alpha_{0}}{\alpha_{0}+\beta_{0}},~~~~\mathcal{M}_{2}=\frac{\alpha_{0}(\alpha_{0}+1)}{(\alpha_{0}+\beta_{0})(\alpha_{0}+\beta_{0}+1)}.
\end{equation}
The parameters $\alpha_{0}$ and $\beta_{0}$ are solved as
in~(\ref{eq:alp0bet0}). This completes the proof.

\section{Exact $T_{\text{ST}}$ Distribution Under $\mathcal{H}_{1}$ for $K=2$}\label{ap:H1K2}

Here we prove Proposition~\ref{th:H1K2}. When $K=2$, the test statistics $T_{\text{ST}}$ reduces to
\begin{equation}
X:=\frac{4\lambda_{1}\lambda_{2}}{(\lambda_{1}+\lambda_{1})^{2}}~~~x\in[0,1].
\end{equation}
Under $\mathcal{H}_{1}$ the joint density of $\lambda_{1}$ and $\lambda_{2}$ is~\cite{1964James}
\begin{equation}
C_{1}\left|\e^{-\frac{\lambda_{1}}{\sigma_{1}}-\frac{\lambda_{2}}{\sigma_{2}}}-\e^{-\frac{\lambda_{1}}{\sigma_{2}}-\frac{\lambda_{2}}{\sigma_{1}}}\right|(\lambda_{1}\lambda_{2})^{N-2}(\lambda_{1}-\lambda_{2})
\end{equation}
where $0\leq\lambda_{2}\leq\lambda_{1}\leq\infty$ and $C_{1}=\frac{(\sigma_{1}\sigma_{2})^{-N-1}}{\Gamma(N-1)\Gamma(N)(\sigma_{2}-\sigma_{1})}$.
Making a change of variables $\lambda_{1},\lambda_{2}$ to $z=\lambda_{1},x=\frac{4\lambda_{1}\lambda_{2}}{(\lambda_{1}+\lambda_{1})^{2}}$ with Jacobian $J=\frac{z(1-\sqrt{1-x})^{2}}{x^{2}\sqrt{1-x}}$ and integrating $z$ out, the density of $X$ reads
\begin{eqnarray}\label{eq:DenH1K2}
&&C_{2}x^{N-2}(1-x)^{-1/2}(1-\sqrt{1-x})^{2N-2}(\sqrt{1-x}+x-1)\times\nonumber\\
&&\left((\sigma_{1}x+\sigma_{2}(1-\sqrt{1-x})^{2})^{1-2N}-(\sigma_{2}x+\sigma_{1}(1-\sqrt{1-x})^{2})^{1-2N}\right),
\end{eqnarray}
where $C_{2}=\frac{2(\sigma_{1}\sigma_{2})^{N}}{B(N,N-1)(\sigma_{2}-\sigma_{1})}$.
In order to obtain the CDF of $X$, we first make a change of variable $w=\sqrt{1-x}$ with Jacobian $J=2w$, the density of $W$ becomes
\begin{eqnarray}
&&2(\sigma_{1}+\sigma_{2})^{1-2N}C_{2}w(1-w)^{N-2}(1+w)^{N-2}\times\nonumber\\
&&\left(\left(1-\frac{\sigma_{1}-\sigma_{2}}{\sigma_{1}+\sigma_{2}}w\right)^{1-2N}-\left(1+\frac{\sigma_{1}-\sigma_{2}}{\sigma_{1}+\sigma_{2}}w\right)^{1-2N}\right).
\end{eqnarray}
Now expanding $\left(1\pm\frac{\sigma_{1}-\sigma_{2}}{\sigma_{1}+\sigma_{2}}w\right)^{1-2N}$ in power series of $w$ and then integrating term-wise the density of $W$, the CDF of $X$ is simplified to~(\ref{eq:H1K2}) by using the fact that $G_{\text{ST}}(y):=\mathbb{P}(Y<y)=1-\mathbb{P}(W<\sqrt{1-y})$. This completes the proof.

\section{Distribution of $T_{\text{ST}}$ Under $\mathcal{H}_{1}$}\label{ap:H1}

Here we prove Proposition~\ref{th:H1}. We first derive an approximative moments expression of the random variable
$T_{\text{ST}}$. Under $\mathcal{H}_{1}$, the sample covariance
matrix $\mathbf{R}$ follows a correlated complex Wishart
distribution $\mathcal{W}_{K}\left(N,\mathbf{\Sigma}\right)$ with
density function
\begin{equation}
\mathbf{R}\sim
\frac{1}{\Gamma_{K}(N)\left(\det(\mathbf{\Sigma})\right)^{N}}\left(\det(\mathbf{R})\right)^{N-K}\e^{\text{tr}(-\mathbf{\Sigma^{-1}R})}.
\end{equation}
The $n$-th moment of random variable $T_{\text{ST}}$ can be
calculated as
\begin{eqnarray}
\mathbb{E}[x^{n}]&=&\frac{K^{Kn}}{\Gamma_{K}(N)\left(\det(\mathbf{\Sigma})\right)^{N}}\int_{\mathbf{R}\succ0}\left(\det(\mathbf{R})\right)^{N-K+n}\e^{\text{tr}(-\mathbf{\Sigma^{-1}R})}\left(\text{tr}(\mathbf{R})\right)^{-Kn}\mathrm{d}\mathbf{R}\\
&=&\frac{K^{Kn}\Gamma_{K}(N+n)}{\Gamma_{K}(N)\left(\det(\mathbf{\Sigma})\right)^{-n}}\int_{\mathbf{R}\succ0}\frac{\left(\det(\mathbf{R})\right)^{N-K+n}\e^{\text{tr}(-\mathbf{\Sigma^{-1}R})}}{\Gamma_{K}(N+n)\left(\det(\mathbf{\Sigma})\right)^{N+n}}\left(\text{tr}(\mathbf{R})\right)^{-Kn}\mathrm{d}\mathbf{R}\\
&=&\frac{K^{Kn}\Gamma_{K}(N+n)}{\Gamma_{K}(N)\left(\det(\mathbf{\Sigma})\right)^{-n}}\mathbb{E}[\left(\text{tr}(\mathbf{R'})\right)^{-Kn}],
\end{eqnarray}
where the last expectation is with respect to $\mathbf{R'}$
distributed as $\mathcal{W}_{K}\left(N+n,\mathbf{\Sigma}\right)$.
The trace of $\mathbf{R'}$ can be represented as~\cite{2007Forenza}
\begin{equation}
Z:=\text{tr}(\mathbf{R'})=\sum_{i=1}^{K}\sigma_{i}z_{i},
\end{equation}
where the random variables $2z_{i}$s are i.i.d Chi-square distributed
with $2(N+n)$ degrees of freedom. The exact density function for $Z$
is available when no multiplicity of $\sigma_{i}$ exists, i.e.
$\sigma_{i}\neq\sigma_{j}$, $\forall i\neq j$~\cite{1992Mathai}.
This effectively requires that $\mathbf{\Sigma}$ is full rank or,
equivalently, the number of active primary users $P$ is greater or
equal to the sensor size $K$. Due to this limitation, we opt for the
Gamma approximation discussed in~\cite{1968Feiveson}, which is still
valid when multiplicities of $\sigma_{i}$ exist. Specifically, for a
Gamma distribution with density
$\frac{1}{\Gamma(a)b^{a}}~x^{a-1}\e^{-\frac{x}{b}}$, the mean and
variance are $ab$ and $ab^{2}$ respectively. For the random variable
$Z$, its mean equals
\begin{equation}
\mathbb{E}[z]=\sum_{i=1}^{K}\sigma_{i}\mathbb{E}[z_{i}]=(N+n)\sum_{i=1}^{K}\sigma_{i}
\end{equation}
and variance equals
\begin{equation}
\mathbb{V}[z]=\sum_{i=1}^{K}\sigma_{i}^{2}\mathbb{V}[z_{i}]=(N+n)\sum_{i=1}^{K}\sigma_{i}^{2}.
\end{equation}
Fitting the mean and variance of a Gamma random variable to those of
$Z$, we obtain the parameters $a$ and $b$ as in~(\ref{eq:GamPara}).
With this Gamma approximation, the $(-Kn)$-th moment for the trace
of $\mathbf{R'}$ is
\begin{equation}
\mathbb{E}[\left(\text{tr}(\mathbf{R'})\right)^{-Kn}]\approx\frac{b^{-Kn}\Gamma(a-Kn)}{\Gamma(a)}.
\end{equation}
Now the approximate moments of $T_{\text{ST}}$ are
\begin{equation}
\mathbb{E}[x^{n}]\approx\left(\frac{K}{b}\right)^{Kn}\frac{\Gamma(a-Kn)\Gamma_{K}(N+n)\left(\det(\mathbf{\Sigma})\right)^{n}}{\Gamma_{K}(N)\Gamma(a)}:=\mathcal{N}_{n}.
\end{equation}
Similar to the case under $\mathcal{H}_{0}$, for a Beta distribution
with parameters $\alpha_{1}$ and $\beta_{1}$, by matching its first
two moments to $\mathcal{N}_{1}$ and $\mathcal{N}_{2}$ we
obtain~(\ref{eq:alp1bet1}). This completes the proof.

\ifCLASSOPTIONcaptionsoff
\newpage
\fi



\begin{thebibliography}{99}

\bibitem{2008Yonghong}
Y. Zeng, C. L. Koh and Y. C. Liang,
\newblock ``Maximum eigenvalue detection: theory and application,'' {\em IEEE International Conference on Communications,} May 2008.

\bibitem{2009Kritchman}
S. Kritchman and B. Nadler,
\newblock ``Non-parametric detections of the number of signals: hypothesis testing and random matrix theory,'' {\em IEEE Trans. Sig. Proc.,} vol. 57, no. 10, pp. 3930-3941, Oct. 2009.

\bibitem{2010Taherpour}
A. Taherpour, M. N. Kenari and S. Gazor,
\newblock ``Multiple antenna spectrum sensing in cognitive radios,'' {\em IEEE Trans. Wire. Commun.,} vol. 9, no. 2, pp. 814-823, Feb. 2010.

\bibitem{2009Lu}
L. Wei and O. Tirkkonen,
\newblock ``Cooperative spectrum sensing of OFDM signals using largest eigenvalue distributions,'' {\em IEEE International Symposium on
Personal, Indoor and Mobile Radio Communications,} Sep. 2009.

\bibitem{2008bZeng}
Y. Zeng, Y. Liang and R. Zhang,
\newblock ``Blindly combined energy detection for spectrum sensing in cognitive
radio,'' {\em IEEE Sig. Proc. Letters,} vol. 15, pp. 649-652, 2008.

\bibitem{2010Wang}
P. Wang, J. Fang, N. Han and H. Li,
\newblock ``Multiantenna-assisted spectrum sensing for cognitive
radio,'' {\em IEEE Tran. Vehi. Tech.,} vol. 59, no. 4, pp.
1791-1800, May 2010.

\bibitem{2010Bianchi}
P. Bianchi, M. Debbah, M. Maida and J. Najim,
\newblock ``Performance of statistical tests for single-source detection using
random matrix theory,'' {\em IEEE Trans. Inf. Theory.} vol. 57, no. 4, pp. 2400-2419, Apr. 2011.

\bibitem{2011Nadler}
B. Nadler, F. Penna and R. Garello,
\newblock ``Performance of eigenvalue-based signal detectors with known and unknown noise
power,'' {\em IEEE International Conference on Communications,} June 2011.

\bibitem{2008Zeng}
Y. Zeng and Y. C. Liang,
\newblock ``Eigenvalue based spectrum sensing algorithms for cognitive
radio,'' {\em IEEE Tran. Commun.,} vol. 57, no. 6, pp. 1784-1793, Jun. 2009.

\bibitem{2009Federico}
F. Penna, R. Garello and M. A. Spirito,
\newblock ``Cooperative spectrum sensing based on the limiting eigenvalue ratio distribution in Wishart matrices,'' {\em IEEE Comm. Letters,} vol. 13, issue 7, pp. 507-509, Jul. 2009.

\bibitem{2010Federico}
F. Penna and R. Garello,
\newblock ``Theoretical performance analysis of eigenvalue-based detection,''
Available at http://arxiv.org/abs/0907.1523

\bibitem{2010Zhang}
R. Zhang, T. J. Lim, Y. C. Liang and Y. Zeng,
\newblock ``Multi-antenna based spectrum sensing for cognitive radios: a GLRT approach,'' {\em IEEE Tran. Commun.,} vol. 58, no. 1, pp. 84-88,
Jan. 2010.

\bibitem{2003Anderson}
T. W. Anderson,
\newblock {\em An Introduction to Multivariate Statistical
Analysis.} Wiley, 2003.

\bibitem{1982Muirhead}
R. J. Muirhead,
\newblock {\em Aspects of Multivariate Statistical Theory.} New York: Wiley, 1982.

\bibitem{1940Mauchly}
J. W. Mauchly,
\newblock ``Significance test for sphericity of a normal n-variate distribution,'' {\em The Annals of Mathematical Statistics}, vol. 11, no. 2, pp. 204-209, Jun. 1940.

\bibitem{1971John}
S. John,
\newblock ``Some optimal multivariate tests,'' {\em Biometrika}, vol. 58, no. 1, pp. 123-127, Apr. 1971.

\bibitem{1972John}
S. John,
\newblock ``The distribution of a statistic used for testing sphericity of normal distributions,'' {\em Biometrika}, vol. 59, no. 1, pp. 169-173, Apr. 1972.

\bibitem{1972Sugiura}
N. Sugiura,
\newblock ``Locally best invariant test for sphericity and the limiting distributions,'' {\em The Annals of Mathematical Statistics}, vol. 43, no. 4, pp. 1312-1316, Aug. 1972.

\bibitem{1975Nagarsenker}
B. N. Nagarsenker and M. M. Das,
\newblock ``Exact Distribution of sphericity criterion in the complex case and its percentage points,''
{\em Communications in Statistics,} 4(4), pp. 363-374, 1975.

\bibitem{1985Nagar}
D. K. Nagar, S. K. Jain and A. K. Gupta
\newblock ``Distribution of LRC for testing sphericity of a complex multivariate Gaussian model,''
{\em Internat. J. Math. \& Math. Sci.,} vol. 8, no. 3, pp. 555-562, 1985.

\bibitem{1969Consul}
P. C. Consul,
\newblock ``The exact distributions of likelihood criteria for different hypotheses,'' {\em Multivariate Analysis 2.} Academic Press, New York, 1969.

\bibitem{2002Simon}
M. K. Simon,
\newblock {\em Distributions Involving Gaussian Random Variables.} New York: Springer, 2002.

\bibitem{1990Williams}
D. B. Williams and D. H. Johnson,
\newblock ``Using the sphericity test for source detection with narrow-band passive arrays,'' {\em IEEE Transactions
on Acoustics, Speech and Signal Processing,} vol. 38, no. 11, pp. 2008-2014, Nov. 1990.

\bibitem{1964James}
James, A.T.
\newblock ``Distributions of matrix variates and latent roots derived from normal samples,'' {\em Ann. Inst. Statist. Math.,} 35, 475-501, 1964.

\bibitem{1971Pillai}
K. C. S. Pillai and B. N. Nagarsenker,
\newblock ``On the distribution of the sphericity test criterion in classical and complex normal populations having unknown covariance matrices,'' {\em The Annals of Mathematical Statistics}, vol. 42, no. 2, pp. 764-767, Apr. 1971.

\bibitem{1971Khatri}
C. G. Khatri and M. S. Srivastava,
\newblock ``On exact non-null distributions of likelihood ratio criteria for sphericity test and equality of two covariance matrices,'' {\em Sankhya}, vol. 33, no. 2, pp. 201-206, Jun. 1971.

\bibitem{2007Forenza}
A. Forenza, M. R. McKay, A. Pandharipande and R. W. Heath,
\newblock ``Adaptive MIMO transmission for exploting the capacity of spatially correlated
channels,'' {\em IEEE Tran. Vehi. Tech.,} vol. 56, no. 2, pp.
619-630, Mar. 2007.

\bibitem{1992Mathai}
A. M. Mathai and S. B. Provost,
\newblock {\em Quadratic Forms in Random Variables.} New York:
Marcel Dekker, 1992.

\bibitem{1968Feiveson}
A. H. Feiveson and F. C. Delaney,
\newblock ``The distribution and properties of a weighted sum of Chi squares,''
{\em NASA Technical Note,} NASA TN D-4575, May 1968.

\bibitem{1961Hochstadt}
H. Hochstadt,
\newblock {\em Special Functions of Mathematical Physics.} Holt, Rinehart and Winston, New York, 1961.

\bibitem{1993Boik}
R. J. Boik,
\newblock ``Algorithm AS 284: Null distribution of a statistics for testing sphericity and additivity: a Jacobi polynomial expansion,'' {\em Journal of the Royal Statistical Society, series C (Applied Statistics)}, vol. 42, no. 3, pp. 567-576, 1993.

\bibitem{2010Nadler}
B. Nadler,
\newblock ``Nonparametric detection of signals by information theoretic
criteria: performance analysis and an improved estimator,'' {\em
IEEE Trans. Sig. Proc.,} vol. 58, no. 5, pp. 2746-2756 , May 2010.

\bibitem{2003Digham}
F. F. Digham, M. Alouini and M. K. Simon,
\newblock ``On the energy detection of unknown signals over fading channels,'' {\em IEEE International
Conference on Communications,} May 2003.

\bibitem{2005Sahai}
A. Sahai and D. Cabric,
\newblock ``Spectrum sensing: fundamental limits and practical challenges,'' tutorial in {\em IEEE International Symposium on New Frontiers in Dynamic Spectrum Access Network,} Nov. 2005.
Available at http://www.eecs.berkeley.edu/$~$sahai/Presentations/Dyspan$\_$2005$\_$tutorial$\_$part$\_$I.pdf

\bibitem{2008Tandra}
R. Tandra and A. Sahai,
\newblock ``SNR walls for signal detetion,'' {\em IEEE J. Select. Topic in Sig. Proc.,} vol. 2, no. 1, Feb. 2008.

\bibitem{1992Sonnenschein}
A. Sonnenschein and P. M. Fishman,
\newblock ``Radiometric detection of spread spectrum signals in noise of uncertain power,'' {\em IEEE Trans. Aerosp.
Electron. Syst.,} vol. 28, pp. 654-660, July 1992.

\bibitem{1984Grieve}
A. P. Grieve,
\newblock ``Tests of sphericity of normal distributions and the analysis of repeated measures designs,'' {\em Psychometrika,} vol. 49, no. 2, pp. 257-267, Jun. 1984.

\bibitem{1988Chan}
Y. M. Chan and M. S. Srivastava,
\newblock ``Comparison of powers for the sphericity tests using both the asymptotic distribution and the bootstrap method,'' {\em Communications in Statistics: Theory and Methods,} 17(3), pp. 671-690, 1988.

\bibitem{1990Boik}
R. J. Boik,
\newblock ``Inference on covariance matrices under rank restrictions,'' {\em Journal of Multivariate Analysis,} 33, pp. 230-246, 1990.




\end{thebibliography}
\end{document}